\documentclass[useAMS,usenatbib]{mn2e}
\usepackage{rotating}
\usepackage{journals}
\usepackage{graphicx, subfigure}
\usepackage{xspace}
\usepackage{amssymb}

\newcommand\msun {M$_{\odot}$}
\def\approxgt{\ifmmode \rlap{$>$}{}_{{}_{{}_{\textstyle\sim}}} \else%
$\rlap{$>$}{}_{{}_{{}_{\textstyle\sim}}}$\fi} 
\def\approxlt{\ifmmode \rlap{$<$}{}_{{}_{{}_{\textstyle\sim}}} \else%
$\rlap{$<$}{}_{{}_{{}_{\textstyle\sim}}}$\fi}

\def\farcs{\hbox{$.\!\!^{\prime\prime}$}}
\def\degr{\hbox{$^\circ$}}
\def\arcmin{\hbox{$^\prime$}}

\def\xmm{XMM-{\it Newton}\xspace}
\def\chan{{\it Chandra}\xspace}
\def\swift{{\it Swift}\xspace}
\def\rxte{{\it RXTE}\xspace}
\def\integral{{\it INTEGRAL}\xspace}
\def\uno{IGR J16194$-$2810\xspace}
\def\due{IGR J16293$-$4603\xspace}
\def\tre{IGR J16479$-$4514\xspace}
\def\quattro{IGR J16500$-$3307\xspace}
\def\cinque{XTE J1710$-$281\xspace}
\def\sei{XTE J1716$-$389\xspace}
\def\sette{IGR J17254$-$3257\xspace}
\def\otto{XTE J1743$-$363\xspace}
\def\nove{IGR J17597$-$220\xspace}
\def\dieci{IGR J18490$-$0000\xspace}
\def\undici{IGR J19308+0530\xspace}

\def\ra{$\rmn{RA}$}
\def\dec{$\rmn{Dec.}$}

\def\sec{$\rmn{s}$ }

\LARGE \normalsize \title[Optical counterparts XRBS]{ \chan localisation and optical/NIR follow-up of Galactic X-ray sources}

\author[Ratti et al.]  {E.M.$\,$Ratti$^{1}$\thanks{email: e.m.ratti@sron.nl},$\,$C.G.$\,$Bassa$^{2}$,$\,$M.A.P.$\,$Torres$^{3}$,$\,$L.$\,$Kuiper$^{1}$,$\,$J.C.A.$\,$Miller-Jones$^{4}$,$\,$P.G.$\,$Jonker$^{1,3,5}$ \\ 
$^1$SRON, Netherlands Institute for Space Research,
Sorbonnelaan 2, 3584~CA, Utrecht, The Netherlands\\ 
$^2$Jodrell Bank Centre for Astrophysics, School of Physics and Astronomy, University of Manchester, Manchester M13 9PL \\  
$^3$Harvard--Smithsonian Center for Astrophysics, 60 Garden Street, Cambridge, MA~02138, U.S.A.\\
$^4$NRAO Headquarters, 520 Edgemont Road,Charlottesville,VA 22903, USA \\
$^5$Department of Astrophysics, IMAPP, Radboud University Nijmegen, PO Box 9010, NL-6500 GL Nijmegen, the Netherlands \\ }

\begin{document}

\maketitle

\begin{abstract} 
\noindent We investigate a sample of eleven Galactic X-ray sources recently discovered with \integral or \rxte with the goal of identifying their optical and/or near-infrared (NIR) counterpart. For this purpose new \chan positions of nine objects are presented together with follow-up observations of all the targets in the optical and NIR. For the four sources  \uno, \tre, \quattro and \undici,  the \chan position confirms an existing association with an optical/NIR object, while for two sources  (\sei and \dieci) it rules out previously  proposed counterparts indicating new ones. In the case of \nove, a counterpart is selected out of the several possibilities proposed in the literature and we present the first association with an optical/NIR source for \due and \otto. Moreover, optical/NIR observations are reported for \cinque and \sette: we investigate the counterpart to the X-ray sources based on their \xmm positions. We discuss the nature of each system considering its optical/NIR and X-ray properties.
  
\end{abstract}


\section{Introduction} 
\label{intro}
\begin{table*}
  \caption{\chan observations. Source counts and positions are given in the table. The positional uncertainty is 0.6 arcsec on all the positions (see Section \ref{Xray}). }
  \label{chandra}
  \begin{center}
  \begin{tabular}{llllllll}
  \hline
  Source & Date & Instrument & Exposure & Counts & \ra(J2000) &
  \dec(J2000) & {\sc wavdetect} error \\
   & & & time (s) & & & & on \ra,~\dec (arcsec) \\
  \hline 
  \uno & 2008~Jan.~18 & {\it HRC-I} & 1129 & 1075 & $16^{\rmn{h}}
  19^{\rmn{m}} 33\fs30$ &  $-28\degr 07\arcmin 40\farcs 30$ & 0.018,~0.014  \\
  \due & 2008~Jan.~24 & {\it ACIS-I} & 1135 & 237 & $16^{\rmn{h}}
  29^{\rmn{m}} 12\fs86$ &  $-46\degr 02\arcmin 50\farcs 94$ &  0.072,~0.068 \\
  \tre & 2007~Oct.~24 & {\it HRC-I} & 1174 & 44 & $16^{\rmn{h}}
  48^{\rmn{m}} 06\fs58$ &  $-45\degr 12\arcmin 06\farcs 74$ & 0.061,~0.061 \\
  \quattro & 2007~Sep.~29 & {\it HRC-I} & 1150 & 198 & $16^{\rmn{h}}
  49^{\rmn{m}} 55\fs65$ &  $-33\degr 07\arcmin 02\farcs 28$ & 0.032,~0.029 \\ 
  \sei & 2008~Sep.~23 & {\it ACIS-I} & 1141 & 12 & $17^{\rmn{h}}
  15^{\rmn{m}} 56\fs42$ &  $-38\degr 51\arcmin 54\farcs 13$ & 0.227,~0.256\\ 
  \otto & 2009~Feb.~08 & {\it HRC-I} & 1172 & 11 & $17^{\rmn{h}}
  43^{\rmn{m}} 01\fs31$ &  $-36\degr 22\arcmin 22\farcs 00$ & 0.14,~0.043 \\ 
  \nove & 2007~Oct.~23 & {\it HRC-I} & 1180 & 227 & $17^{\rmn{h}}
  59^{\rmn{m}} 45\fs52$ &  $-22\degr 01\arcmin 39\farcs 17$ & 0.022,~0.022 \\ 
  \dieci & 2008~Feb.~16 & {\it HRC-I} & 1174 & 22 & $18^{\rmn{h}}
  49^{\rmn{m}} 01\fs59$ &  $-00\degr 01\arcmin 17\farcs 73$ & 0.0432,~0.061 \\ 
  \undici & 2007~Jul.~30 & {\it HRC-I} & 1129 & 26 & $19^{\rmn{h}}
  30^{\rmn{m}} 50\fs77$ &  $+05\degr 30\arcmin 58\farcs 09$ & 0.061,~0.072 \\ 
  \end{tabular}
  \end{center}
\end{table*}

\begin{table*}
  \caption{Known X-ray  positions}
  \label{known}
  \begin{center}
  \begin{tabular}{lllll}
  \hline
  Source & \ra(J2000) & \dec(J2000) & Uncertainty (1$\sigma$) & Instrument \\
  \hline 
  \cinque&  $17^{\rmn{h}} 10^{\rmn{m}} 13\fs$ &  $-28\degr 07\arcmin 51\farcs $ & $1''$  & \xmm$^{a}$ \\
  \sette & $17^{\rmn{h}} 25^{\rmn{m}} 25\fs$ &  $-32\degr 57\arcmin 15\farcs $ & $2''$ & \xmm$^{b}$ \\
  \end{tabular}
  \end{center} {\footnotesize $^a$ \citealt{2008yCat..34930339W}. $^b$ \citealt{2007A&A...469L..27C}. \\}
\end{table*}

X-ray binaries (XRBs) are binary systems where a compact object, either a black
hole (BH), a neutron star (NS) or a white dwarf (WD) accretes matter from a
stellar companion (the donor or secondary star). In case the compact object is a WD, the XRB is called a cataclysmic variable (CV). In these systems, gravitational potential energy is extracted from the matter falling onto the compact object via the accretion process, producing the observed X-ray luminosity.
XRBs represent a large fraction of X-ray sources in our Galaxy (see \citealt{2006csxs.book....1P} for a review).

The majority of XRBs accreting onto a NS or BH can be grouped in two classes, defined by the mass of the secondary star: high mass X-ray
binaries (HMXBs) and low mass X-ray binaries (LMXBs). In HMXBs the mass
of the donor is M$_\rmn{D}\gtrsim10$ \msun; in LMXBs M$_\rmn{D}\lesssim1$ \msun. A
few intermediate mass XRBs (IMXBs) are also known (see \citet{2003astro.ph..8020C} for a review).

HMXBs are further divided into Be-XRBs and supergiant X-ray
binaries (SXRBs). Be-XRBs are characterized by a Be-star companion and typically have eccentric orbits: the compact object accretes in major outbursts near the periastron, when it passes through the
circumstellar disk of the Be-star. On the other hand, SXRBs host  an early type O/B supergiant companion and are 'traditionally' found to be persistent X-ray sources. However, recent
observations by the International Gamma-ray Astrophysics Laboratory (\integral) have revealed a class of fast X-ray
transient sources spending most of their time at a quiescent level, that have sporadic outbursts lasting a few minutes to
hours. The class was named "supergiant fast X-ray transients" after follow-up optical and near-infrared (NIR) spectroscopic observations of a number of systems revealed supergiant secondary stars \citep{2006ESASP.604..165N}. The physical origin of the fast X-ray
outbursts is not yet understood (for different models see \citealt{2005A&A...441L...1I}, \citealt{2007A&A...476.1307S}, \citealt{2008MNRAS.391L.108B}, \citealt{2010arXiv1006.3256D}). 
The \integral satellite also 
discovered a population of XRBs characterized by a large amount of absorption
local to the source \citep{2005A&A...444..821L} as the accreting compact object is immersed in the dense stellar wind of a
massive companion star. The NIR counterparts identified for
a number of sources are all consistent with
supergiant stars. It has been suggested that all the obscured HMXBs are
hosting supergiant companions \citep{2006A&A...453..133W}.

LMXBs are traditionally divided into NS and BH binaries. Surface phenomena occurring on the accreting object, like thermonuclear X-ray bursts or the detection of a pulsating signal
are evidence for the presence of a NS. Nevertheless, in the absence of such
phenomena no definitive conclusion can be drawn about the nature of the compact object from X-ray observations
alone (see \citealt{2006csxs.book....1P} for a review). Dynamical constraints on the mass of the compact object are required in order to confidently distinguish a NS from a BH. These can be obtained via
orbital phase-resolved spectroscopy of the optical or NIR counterpart
to the X-ray source \citep{1995xrbi.nasa...58V}. 

Two sub-classes of LMXBs also exist that are characterised by peculiar companion stars: ultra compact X-ray binaries
(UCXBs) and symbiotic X-ray binaries (SyXBs). The signature of UCXBs
is an orbital period of $\sim$ 1 hour or less. This implies that the orbital
separation is very small and the donor  must be hydrogen poor
to fit in its Roche lobe \citep{2007A&A...465..953I}. SyXBs are defined by the presence of an M-type giant companion. Those systems are rare, as the giant
phase does not last long, and are characterised by a lack of accretion signatures in the optical spectra, since the
accretion disk is out-shined by the companion star unless the X-ray
luminosity is particularly high (\citealt{2007A&A...470..331M} and
references therein).

Since the classification is based on the mass of the companion and/or of the compact object, the identification of counterparts of XRBs in optical or NIR is important. In this article we present a search for the optical/NIR
counterpart of a sample of 11 Galactic sources, recently discovered
with \integral or with the Rossi X-ray Timing Explorer ({\it RXTE}).
A classification has been proposed in the literature for all but one
of the sources (IGR J16293-4603) on the basis of the X-ray
behaviour alone or the spectrum of an optical/NIR candidate counterpart. Nevertheless, none of the proposed
counterpart identifications are conclusive due to the lack of an accurate X-ray
position.  For the 9 sources listed in Table~\ref{chandra} we obtained \chan
observations, with the main goal of determining an accurate X-ray position. Thanks to the high spatial resolution of \chan we
could verify previously proposed candidate counterparts and
investigate sources for which no counterpart was known.  For
the two sources listed in Table~\ref{known} we do not have
\chan data. We have searched for their optical/NIR counterparts referring to previous \xmm observations. 

This paper is structured as follows: sections 2 and 3 present the observations and the data reduction procedures. In section 4 we provide a short introduction for each source followed by the results from our analysis.
All the coordinates reported in the text and tables are referred to epoch J2000.

\section{X-ray data: reduction and analysis} 
\label{Xray}

We observed the sources in Table~\ref{chandra} with \chan, using the High Resolution Camera ({\it HRC-I}) and the Advanced CCD Imaging Spectrometer ({\it ACIS-I}). 
We have reprocessed and analysed the data using the {\sc ciao} 4.0.1 software developed by
the \chan X-ray Centre. All data have been used in our analysis, as background flaring is very weak or absent.

\noindent We localized X-ray sources on each observation with the tool {\sc wavdetect} from the total energy range of {\it HRC-I} and {\it ACIS-I}. The uncertainty on the localisation on the image as given by {\sc wavdetect} (Table \ref{chandra}) is negligible with respect to the \chan boresight uncertainty of 0.6 arcsec (90 per cent confidence, slightly dependent on the
instrument\footnote{http://cxc.harvard.edu/cal/ASPECT/celmon/}) for all the sources but the weakest, \sei and \otto. Although the centroiding uncertainty for those targets is of the same order of magnitude as the boresight uncertainty, the latter still dominates the overall X-ray positional accuracy. Therefore we adopt a 90 per cent confidence uncertainty of 0.6 arcsec on the X-ray position of all the sources. We extracted the source counts in a 40-pixel radius around the position from {\sc wavdetect} using the tool {\sc dmextract}. We estimated the background in an annulus centered on the {\sc wavdetect} position, with an inner and outer radius of 70 and 200 pixels. We considered all the counts from {\it HRC}, while we select the counts in the $0.3-7~\rmn{keV}$ energy band for {\it ACIS} observations. 
The net, background subtracted counts for each source are given in Table~\ref{chandra}.

\section{Optical and NIR data: reduction and analysis} 

\label{optical}

\begin{table*}
\caption{Properties of the instruments employed for optical/NIR observations.}
\begin{center}
\label{instruments}
\begin{tabular}{lllll}
\hline
Instrument   & Pixel scale            &    Binning &    Field of view  \\
             & ($\rmn{arcsec}$) & & ($\rmn{arcmin}$) \\
\hline
{\it EMMI}   & 0.166 & 2$\times$2 & 9.1$\times$9.9 \\
{\it IMACS}  & 0.111 & 2$\times$2 & 15.4$\times$15.4 \\
{\it LDSS3}  & 0.189 & 1$\times$1 & Diameter$=$8.3 \\
{\it MOSAIC} & 0.27 & 1$\times$1 & 36$\times$36 \\
{\it PANIC}  & 0.127 & 1$\times$1 & 2$\times$2 \\

\end{tabular}
\end{center}
\end{table*}

\begin{table*}
  \caption{Journal of the optical/NIR observations.}
  \label{log}
  \begin{center}
  \begin{tabular}{lllllll}
 \hline
  Source & Date & Instrument(s) & Filter(s) & Exposures  & Seeing & Photometric\\
         &      &               &           &            & ($\rmn{arcsec}$) & calibration\\
  \hline
  \uno&2006\,Jun.\,24 & {\it MOSAIC\,II} & {\it r'} & 1$\times$10\sec+5$\times$300\sec & 1.6 &  PG1323-086\\
      &2006\,Aug.\,03 & {\it PANIC} & $K_s$ & 1$\times$75\sec & 0.5 & 2MASS\\
  \due & 2008\,Jun.\,24 & {\it LDSS3} & {\it g',\,r',~i'} & 3$\times$180\sec\,({\it r'})\,+\,1$\times$180\sec\,({\it g',i'}) & 1.1 & SDSS$^{(1)}$(s82) \\  
  \tre & 2006\,Jun.\,24 & {\it MOSAIC~II} & {\it i'} & 1$\times$10\sec+5$\times$300\sec & 1.6 &  PG1323-086\\
  \quattro & 2006\,Jun.\,24 & {\it MOSAIC\,II} & {\it r'} & 1$\times$10\sec+5$\times$300\sec & 1.6 &  PG1323-086\\
           &  2006\,Aug.\,03 & {\it PANIC} & $K_s$ & 1$\times$75\sec & 0.7 & 2MASS\\
  \cinque & 2005\,May\,07 & {\it IMACS} & {\it I} & 2$\times$10\sec+2$\times$300\sec & 0.9 & Landolt109-954\\
  \sei & 2009\,May\,07 & {\it LDSS3} & {\it i'} & 2$\times$10$\rmn{s}$+4$\times$300\sec & 0.75 & non-photometric$^{(2)}$\\
  \sette & 2006\,Aug.\,03 & {\it PANIC} & {\it $K_s$} & 1$\times$75\sec & 0.8 & 2MASS\\
  \otto & 2007\,Jun.\,22 & {\it EMMI} & {\it I} & 1$\times$30\sec+1$\times$600\sec & 1.7 & non-photometric\\
  \nove & 2009\,May.\,07 & {\it LDSS3} & {\it i'} & 2$\times$10\sec+4$\times$300\sec & 0.75 & non-photometric$^{(2)}$ \\ 
        & 2007\,Jun.\,22 & {\it EMMI} & {\it I} & 1$\times$20\sec+2$\times$600\sec & 1.7 & non-photometric \\
  \dieci & 2006\,Jun.\,25 & {\it MOSAIC\,II} & {\it i'} & 1$\times$10\sec+5$\times$300\sec & 1.1 &  PG1323-086\\
         &   2009\,Jul.\,16 & {\it PANIC} & $K_s$ & 1$\times$450\sec & 0.8 & 2MASS \\
\undici & 2006\,Jun.\,22 & {\it MOSAIC\,II} & {\it r'} & 1$\times$10\sec+5$\times$300\sec & 1.1 &  PG1323-086\\
  \end{tabular}
  \end{center} {\footnotesize $^{(1)}$ SDSS$=$Sloan Digital Sky Survey $^{(2)}$ see text (Section \ref{optical}) \\}
\end{table*}

We performed optical and/or NIR imaging of the field of each X-ray source in our sample from various Chilean sites, with the following instruments: 
\begin{itemize}
\item[-]the ESO Multi Mode Instrument ({\it EMMI}) at the 3.5 m New Technology Telescope on La Silla
\item[-]the Low Dispersion Survey Spectrograph ({\it LDSS3}), the Persson's Auxiliary Nasmyth Infrared Camera ({\it PANIC}) and the Inamori Magellan Areal Camera and Spectrograph ({\it IMACS}) at the 6.5 m Magellan telescopes Clay and Baade on Las Campanas
\item[-]the {\it MOSAIC~II} imager at the 4 m Blanco telescope on Cerro Tololo. 
\end{itemize}
Table \ref{instruments} reports the pixel scale and the field of view (FOV) of each instrument, together with the binning we employed.
A journal of the observations is presented in Table~\ref{log}. All optical observations include a short (10-15 $\rmn{s}$) exposure image for the astrometry, where bright stars do not saturate, and several deeper exposures to observe faint objects. We observed four of the sources in the $K_s$ band using the PANIC camera. The observations consisted of five point dither patterns with a 5s or 15s exposure repeated three times at each offset position. Table ~\ref{log} gives the total time expended on source. 

Optical images have been reduced for photometry with standard routines running within {\sc midas} or {\sc iraf}, corrected for the bias and flat-fielded. The {\it PANIC} NIR data were reduced through the {\it PANIC} software: the raw frames were first dark subtracted and flat-fielded. Normalized flat-fields were made by combining twilight flat field frames scaled by their mode. Next, a sky image was built by masking out stars from each set of dithered frames and was subtracted from the set of target frames. Finally, a mosaic image was built by combining and averaging the sky-subtracted images. 

The {\sc daophot~II} package \citep{1987PASP...99..191S}, running inside {\sc midas}, was used to determine instrumental magnitudes through a Point Spread Function (PSF) fitting technique. The aperture correction was measured from aperture photometry on bright and isolated stars. Unless we detect the counterpart to an X-ray source only on deep images, we preferably perform photometry on the short-exposure astrometric images, which have the advantage of being less crowded than deeper ones.

 The last column in Table \ref{log} lists the fields employed for photometric calibration. The {\it PANIC}  {\it K}$_{s}$-band images have been calibrated with respect to {\it K} band magnitudes of 2MASS stars in the field. An accurate calibration for the LDSS3 observations of  \sei and \nove is not possible since the observing night was not photometric. Nevertheless we provide indicative magnitudes by calibrating the images with the zero point from a previous {\it LDSS3} observing run in November 2007, correcting for the different air mass.

For the astrometry, we compared the position of the stars against entries from the second USNO CCD Astrograph Catalogue (UCAC2, \citealt{2004AJ....127.3043Z}) or from the Two Micron All Sky Survey (2MASS). The positional accuracy of UCAC2 varies between 0.02 arcsec (for stars with magnitude $R<14$) and 0.07 arcsec ($14<R<16$); that of 2MASS is $\sim$0.1 arcsec  (for stars with magnitude $K<14$). UCAC2 positions were preferably adopted, unless less than 5 stars from that catalogue overlap with stars in the field. In this case we compared with the more rich but less precise 2MASS catalogue.  An astrometric solution was computed by fitting for the reference point position, the scale and the position angle, considering all the stars that are not saturated and appear stellar and unblended. We obtain solutions with root-mean-square (rms) residuals ranging from 0.05 to 0.1 arcsec when 2MASS is used, and from 0.05 to 0.07 arcsec in case UCAC2 is used. The uncertainty on the position of a star due to centroiding is negligible with respect to that of the astrometry. 
Once astrometrically calibrated, short-exposure images  have been adopted as secondary catalogues for the calibration of the longer-exposure images (see Table \ref{log}), obtaining rms residuals negligible with respect to those of the 'primary' solution against the standard catalogues. We adopted as the accuracy on our stellar positions the quadratic sum of the residuals of the 'primary' astrometry and the accuracy of the catalogue employed (although the latter could be a systematic error) : the resulting positional accuracy is ranging from 0.07 to 0.13 arcsec (1$\sigma$) on both right ascension (\ra) and declination (\dec). In order to identify the optical/NIR counterpart of the X-ray sources in our sample, we plotted the 90 per cent confidence error circle around \chan or \xmm positions on optical/NIR charts, taking into account the positional error due to our astrometry. For all the \chan targets we searched for a counterpart inside an overall 90 per cent confidence radius of $\sim$0.7 arcsec, resulting  from the combination of the 0.6 arcsec accuracy of \chan positions with the accuracy of our astrometry. The contribution of the astrometric error to the \xmm positional error (see Table \ref{known}) is negligible.

When the candidate counterpart is not saturated, we compute the probability that it falls inside the \chan error circle by chance as the number of stars of brightness equal to or larger than that of the counterpart (considering the error on the photometry), divided by the area of the field and multiplied by the area covered by the \chan error circle.  For stars that saturate even in a 10-15 seconds image we do not compute a probability, since their magnitude cannot be reliably determined.

\section{Individual sources} 
 \label{indsou}
\subsection{IGR J16194-2810: a SyXB}
\label{uno}

\uno was discovered by {\it INTEGRAL/IBIS} (\citealt{2006ApJ...636..765B}; \citealt {2006ApJ...636L..65B}) and soon identified as the ROSAT object 1RXS J161933.0-280736 \citep{2006A&A...445..869S}.  Based on its \swift{\it /XRT} and {\it ROSAT} position, \citet{2007A&A...470..331M} associated the source with the bright object USNO-A2.0~U0600\_20227091. The optical spectrum presented by the authors indicates an M2~III star, thus \uno was classified as a SyXB \citep{2007A&A...470..331M}. 

We observed the field with {\it Chandra/HRC-I} for $\sim1.1~\rmn{ks}$ on 2008 Jan. 18, detecting a single, bright source (1075 counts) inside the {\it ROSAT} and \swift error circles at \ra=~$16^{\rmn{h}}$~$19^{\rmn{m}}$~$33\fs42$, \dec=~$-28\degr$~$07\arcmin$~$40\farcs 3$.

 NIR and optical images, collected with {\it Magellan/PANIC} in $K_{s}$ band on 2006~Aug.~3 and with {\it Blanco/MOSAIC~II} in {\it r'} band on 2006~Jun.~24, reveal a bright source overlapping with the \chan error circle. The source is not saturated only in the 10$\rmn{s}$-long exposure with {\it MOSAIC~II} shown in Figure \ref{first}. It falls on the border of the 90 per cent \chan error circle, at  \ra=~$16^{\rmn{h}}$~$19^{\rmn{m}}$~$33\fs346$, \dec=~$-28\degr$~$07\arcmin$~$39\farcs 92$ ($\pm$0.08 arcsec on both coordinates). Comparing our finding charts with those in \citet{2007A&A...470..331M} we identify this object with the candidate counterpart proposed by the authors. The position reported in the USNO-A2.0 catalogue from observations performed in 1979 is in agreement with our measurements if the strong proper motion of the source is taken into account ($1.3~\pm~4.7~\rmn{mars\,yr^{-1}}$ in \ra$~$and  $-20.2~\pm~4.7~\rmn{mars\,yr^{-1}}$ in \dec, from UCAC2). The magnitudes of the object from USNO-A2.0 in the {\it R} and {\it B} bands is $R=11$ and $B=13.2$. We found the source also in UCAC2 and in 2MASS, where the following magnitude are reported:  $J=8.268~\pm~0.029$, $H=7.333~\pm~0.044$ and $K=6.984~\pm0.016$. We measure an apparent magnitude ${r'}=10.98~\pm~0.04$ in {\it r'} band. 
The probability that such a bright star falls by chance in the Chandra error-circle is $\sim2\times10^{-7}$. Based on its position and proper motion we confirm the red giant USNO-A2.0~U0600\_20227091/2MASS~16193334-2807397 (first proposed by \citealt{2007A&A...470..331M})  as the optical and NIR counterpart to \uno and the classification of the source as a SyXB.

\begin{figure}
\includegraphics[width=9cm, angle=0]{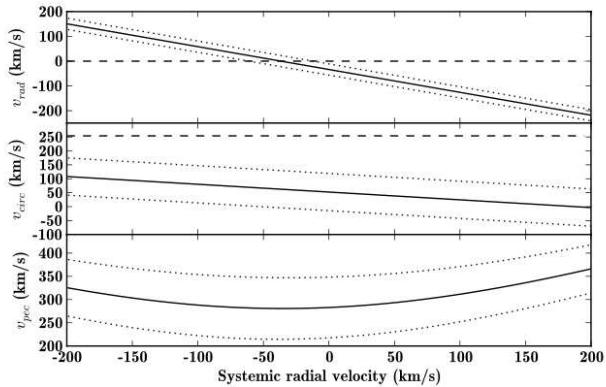}
\caption{From top to bottom, modelled Galactocentric radial velocity and circular velocity and peculiar velocity  of \uno against systemic radial velocity:  the solid line in each case shows the best fitting values while the dotted lines show the uncertainty. The dashed lines show the expected values for Galactocentric
radial and circular velocities for an object participating in the Galactic
rotation. The minimum peculiar velocity is 280$\pm66~\rmn{km\,s^{-1}}$, corresponding to
a systemic radial velocity of $-35~\rmn{km\,s^{-1}}$. }
\label{veloc}
\end{figure}

 We investigated the high proper motion of the source by calculating its peculiar velocity, i.e.\ the velocity with respect to the local standard of rest.  The distance $d$ can be estimated by comparing the apparent magnitude we measure in the {\it r'} band with the typical {\it R}-band absolute magnitude of an M2 III star ( $M_R\sim-2$, \citealt{2000asqu.book.....C}), accounting for extinction. We obtain the extinction coefficient in the {\it R} band from that in the {\it V} band following the optical extinction laws in \citet{1989ApJ...345..245C}. The standard value of the extinction-law parameter $R_V$ for the diffuse interstellar medium is assumed ($R_V=3.1$). We derived $A_V$ from the hydrogen column density $N_H$ in accordance with \citet{2009MNRAS.tmp.1480G}. With $N_H=(0.16\pm0.08)\times10^{22}~\rmn{cm}^{-2}$ from \citet{2007A&A...470..331M}, the distance is $d=3.0\pm0.2~\rmn{kpc}$, in agreement with the upper limit estimated by those authors.  While the systemic radial velocity, $\gamma$, is unknown, we can use the measured proper motion and source distance to derive the three-dimensional space velocity components as a function of $\gamma$. 
Using the transformations of \citet{1987AJ.....93..864J} and the standard solar motion of \citet{1998MNRAS.298..387D} we derive the  space velocity components and compare with those predicted by the Galactic rotation parameters of \citet{2009ApJ...693..397R} (but note \citealt{2009MNRAS.400L.103M}), obtaining the peculiar velocity as a function of $\gamma$ (Figure \ref{veloc}) under the asumption that the
object participates in the Galactic rotation. We find a  minimum peculiar velocity of $280\pm66$\,km\,s$^{-1}$, at $\gamma=-35$\,km\,s$^{-1}$. This limit on the peculiar velocity is high, indicating that either the binary is a halo object or that it has received a kick. The latter possibility is more natural in the case of a NS/BH accretor.

\subsection{IGR J16293-4603: a new LMXB (possibly SyXB)}
\label{due}

The source \due was discovered in 2008 combining {\it INTEGRAL IBIS/ISGRI} data collected over the period from 2003~Mar.~2 to 2006~Feb.~24. The discovery is reported in \citet{2008ATel.1774....1K}, together with a \chan localisation and preliminary results regarding the optical counterpart of the source: in this paper we report the conclusive results of that analysis.

We observed the field of \due with {\it Chandra/ACIS-I} for $\sim1.1~\rmn{ks}$ on 2008 Jan. 24, detecting a single source (237 counts) inside the {\it IBIS/ISGRI} error circle, at \ra=~$16^{\rmn{h}}$~$29^{\rmn{m}}$~$12\fs86$, \dec=~$-28\degr$~$02\arcmin$~$50\farcs 94$.

\begin{figure}
\includegraphics[width=6cm,angle=-90]{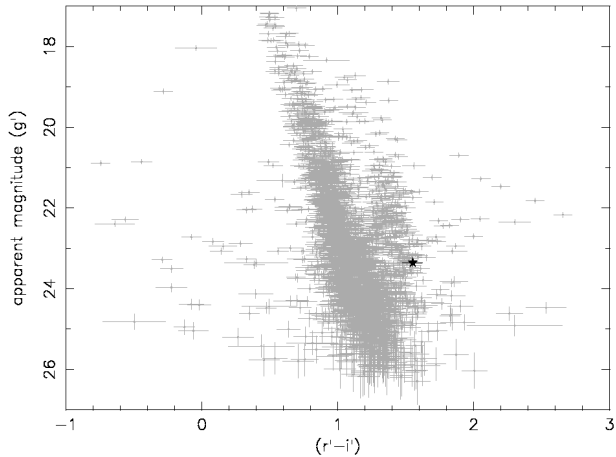}
\caption{ Colour-magnitude diagram of the field of \due, observed with LDSS3. The apparent magnitude of the stars in {\it g'} band is plotted versus the ($r'-i'$) colour index. The counterpart to \due is indicated by the black star. }
\label{hr}
\end{figure}

\begin{figure*}
\begin{minipage}[b]{8.5 cm} 
\centering
\includegraphics[width=5.4 cm]{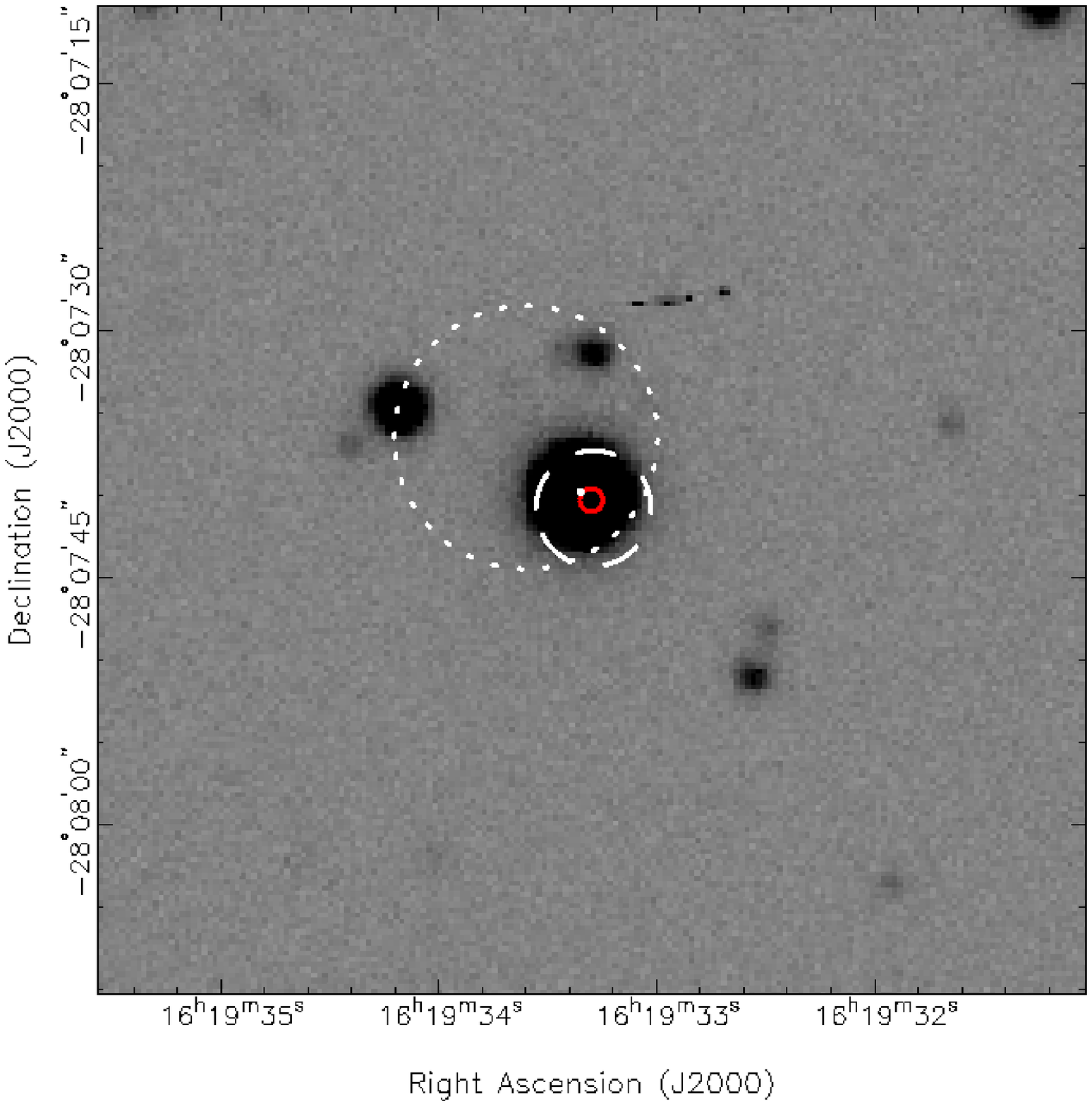}
\caption{\uno: {\it MOSAIC~II}, 300 $\rmn{s}$ in the {\it r'} band. The red error circle indicates the \chan position. The small-dashed error circle is that of {\it ROSAT}, the large-dashed one is \swift.}
\label{first}
\end{minipage}
\hspace{0.5 cm}
\begin{minipage}[b]{8.5 cm}         
\centering
\includegraphics[width=5.7 cm]{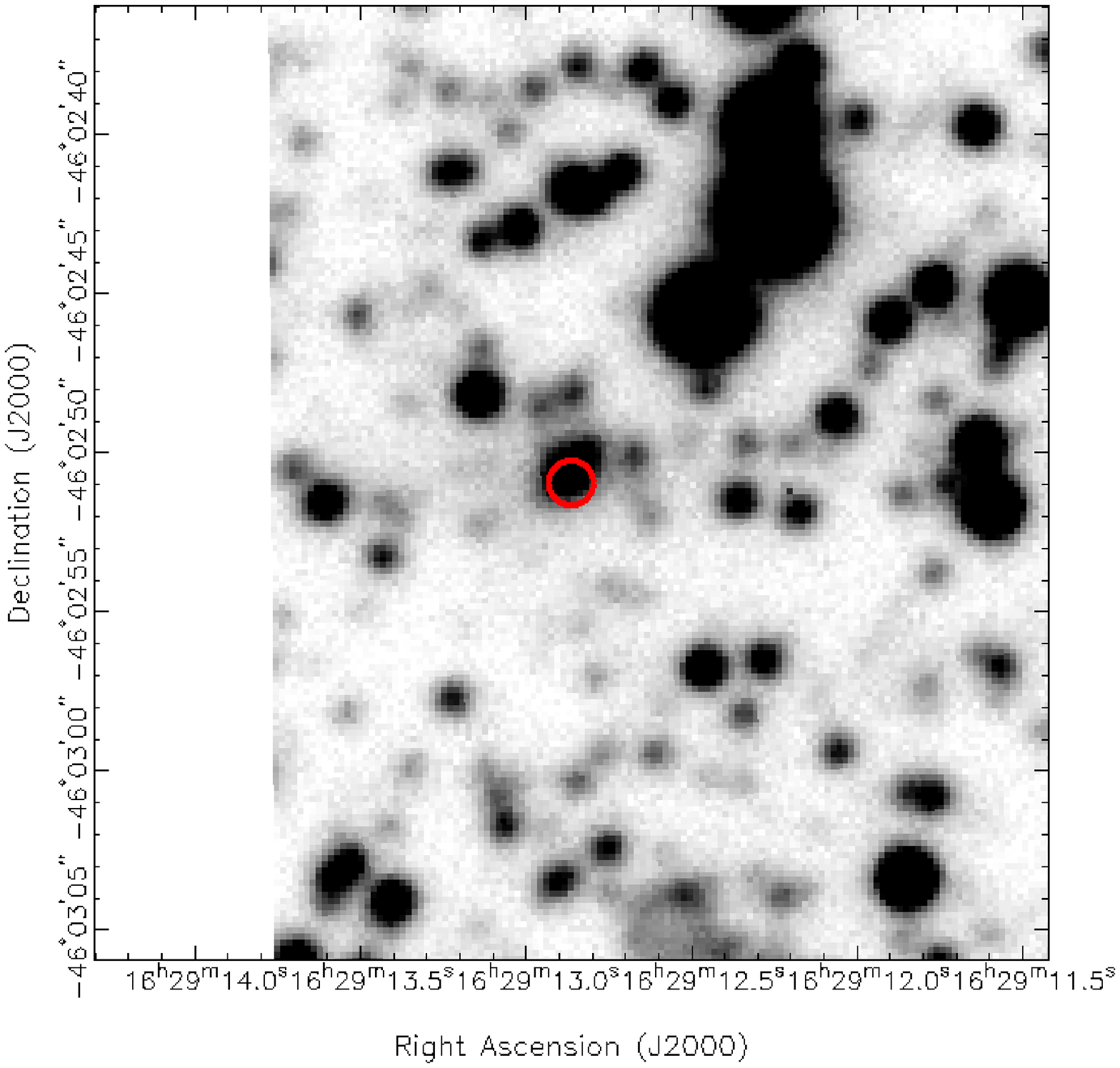}
\caption{\due: {\it LDSS3}, 180 $\rmn{s}$ in the {\it i'} band. The error circle indicates the \chan position.\newline}
\label{second}
\end{minipage}
\end{figure*}

 Optical images have been acquired with {\it Magellan/LDSS3} on 2008~Jun.~24 in the {\it g', r'} and {\it i'} bands. An object is visible in all the observed bands inside the 90 per cent \chan error circle (see Figure \ref{second}) at \ra=~$16^{\rmn{h}}$~$29^{\rmn{m}}$~$12\fs885$, \dec=~$-46\degr$~$02\arcmin$~$50\farcs 55$ ($\pm~0.1$ arcsec on both coordinates). After absolute photometric calibration, we measure the following magnitudes: $g'=23.35~\pm~0.07$, $r'=20.67~\pm~0.04$ and  $i'=19.12~\pm~0.07$. 
 In order to constrain the intrinsic colour index $(r'-i')_0$ for the counterpart to \due, we derived 
the extinction coefficients in the {\it g', r'} and {\it i'} bands as we did for \uno, obtaining the extinction coefficient in the {\it V} band, $A_V$ from the hydrogen column density $N_H$. 
 With  $N_H=(0.7~\pm~0.5)\times10^{22}~\rmn{cm^{-2}}$, as measured in \citet{2008ATel.1774....1K} from the fitting of \chan-{\it ACIS} spectra in the 0.3-7 $\rmn{keV}$ range, we obtain $A_V=3~\pm~2$ and $(r'-i')_0=0.9~\pm~0.4$. If the counterpart we observe has no flux contribution from an accretion disk, this colour index indicates a main sequence star of K or early M spectral type or a giant \citep {2000asqu.book.....C}. If we are observing a combination of the optical light emitted by the disc and by the companion star, the latter is even redder than $(r'-i')_0=0.9~\pm~0.4$, since the disk is bluer than a  K-type star \citep{1995xrb..book...58V}.  Therefore,  \due is most likely not a HMXB. Single stars have been observed in hard X-rays only during flares \citep{2007ApJ...654.1052O}: the fact that \due was discovered by combining multiple \integral observations suggests that the X-rays are not due to a single active star. Thus, \due is most likely an LMXB or a CV. Moreover, Figure \ref{hr} shows the colour-magnitude diagram of the source field, where the apparent magnitude of the stars in {\it g'} band is plotted versus the $(r'-i')$ colour index: the counterpart to \due lies on the Giant Branch of the diagram, suggesting the system has a giant companion. For $N_H\sim0.7\times10^{22}$, the source has the $(r'-i')_0$ of a K5-M0 giant. This value of $N_H$ is in the middle of the range allowed by \chan measurements and corresponds to the Galactic value from \citet{1990ARA&A..28..215D}. For extreme values  of the $N_H$ within the error of \integral measurements, the companion could also be a G5-M2 giant ($N_H$ respectively lower or higher than the Galactic value). The typical absolute magnitude in {\it R} band of an M2 giant is $M_R\sim-1.94$; that of a G5 is $M_R\sim0.2$  \citep{2000asqu.book.....C}. Comparing these values with the magnitude we observe in the {\it r'}-band and assuming the appropriate column density in the two cases, we can estimate the distance $d$ to the source:
$d\sim28~\rmn{kpc}$ if the donor is an {\it M2-III} type and $d\sim45~\rmn{kpc}$ if the donor is a {\it G5-III} type. The X-ray flux from our \chan observations is $4\times10^{-12}~\rmn{erg~s^{-1}~cm^{-2}}$ in the band 0.3-7 $\rmn{keV}$, assuming a simple power-law spectrum with photon index $\gamma=$1.0 \citep{2008ATel.1774....1K}. This results in a luminosity of $\sim4\times10^{35}~\rmn{erg~s^{-1}}$ at 28 $\rmn{kpc}$ and  $\sim2\times10^{36}~\rmn{erg~s^{-1}}$ at $45~\rmn{kpc}$. Intermediate luminosities and distances are obtained for a  K-type companion. All the possibilities lead to an X-ray source that is too bright for a CV, but consistent with an LMXB (although in the case of a  G secondary star the source would be located very far in the halo). We conclude that \due is an LMXB, probably with a giant companion of spectral type  K,  M or, less likely, a  G. Since the donor can also be an  M-type giant, we also indicate \due as a candidate SyXB.

\subsection{IGR J16479-4514: an eclipsing SFXT}
\label{tre}

\tre was discovered with the {\it IBIS/ISGRI} detector on board the \integral observatory on 2003~Aug.~8-9 \citep{2003ATel..176....1M} and observed several times by the same satellite during the following years (\citealt{2005A&A...444..221S}; \citealt{2006ATel..816....1M}). It has been regularly monitored with \swift from October 2007 to October 2008 (\citealt{2008A&A...487..619S}; \citealt{2008ApJ...680L.137R} and \citealt{2009MNRAS.399.2021R}) and observed with \xmm in 2008 \citep{2008MNRAS.391L.108B}. The X-ray behaviour of the source is typical of SFXTs, characterized by short outbursts that have been observed with both \integral and \swift (\citealt{2005ATel..599....1K}; \citealt{2006ApJ...646..452S}; \citealt{2007A&A...476..335W}; \citealt{2009ApJ...690..120S}). Evidence of possible X-ray eclipses is presented in \citet{2008MNRAS.391L.108B} on the basis of \xmm observations and has been recently confirmed by \citet{2009MNRAS.399.2021R} from the analysis of {\it Swift/BAT} data. The orbital period obtained from the eclipses is $\sim3.3$ days, short compared to other SFXTs. Moreover, the luminosity of \tre is $\sim10^{34}~\rmn{erg~s^{-1}}$ in quiescence. This is the typical luminosity of the fainter persistent SXRBs and two orders of magnitude higher than typical for SFXTs. This suggests that \tre is persistently accreting at a low level, in agreement  with its short orbital period. Due to its quiescent luminosity level, compatible with 'canonical' SXRBs, combined with the short outbursts typical of SFXTs, \tre has been proposed as the missing link between the two classes (i.e. \citealt{2009MNRAS.397L..11J}). 

The 2MASS star J16480656-4512068 has been proposed as a possible counterpart to \tre by \citet{2005ATel..599....1K} and \citet{2006A&A...453..133W}.  NIR spectra of that object are presented in \citet{2008A&A...484..783C} and \citet{2008A&A...486..911N}, indicating an O/B supergiant. This is supported by the SED in \citet{2008A&A...484..801R}. In particular, \citet{2008A&A...486..911N}   classify the source as a spectral type O9.5 Iab. A second, fainter candidate counterpart  is also indicated in \citet{2008A&A...484..783C} in {\it K} band, inside the 4 arcsec \xmm error circle. 

In order to select the actual counterpart to \tre, we observed the field with {\it Chandra/HRC-I} for $\sim1.2~\rmn{ks}$ on 2007~Oct.~24.  A single source (44 counts) is detected in the \xmm error circle, at coordinates \ra=~$16^{\rmn{h}}$~$48^{\rmn{m}}$~$06\fs6$, \dec=~$-45\degr$~$12\arcmin$~$06\farcs 7$ 

 Follow-up observations,  performed with {\it Blanco/MOSAIC~II} in the {\it i'} band on 2006 June 24, revealed no candidate counterpart inside the \chan 90 per cent confidence error circle , down to a limiting magnitude of $i'\sim23$. We detect the object labelled {\it 2} in \citet{2008A&A...484..783C} at \ra=~$16^{\rmn{h}}$~$48^{\rmn{m}}$~$06\fs56$, \dec=~$-45\degr$~$12\arcmin$~$08\farcs 1$ ($\pm~0.1$ arcsec on both coordinates) inside the \xmm error circle but outside the \chan one (see Figure \ref{fifth}). We can exclude that source as a counterpart to IGR J16479-4514. We do not detect the candidate counterpart labeled  {\it 1} in \citet{2008A&A...484..783C} in {\it i'} band, but this is not surprising on the basis of its NIR spectrum. In order to verify its association with \tre we compared its coordinates from 2MASS with our \chan position, finding a separation of 0.2 arcsec ($< 1\sigma$). Based on its position, we confirm the object 2MASS~J16480656-4512068  (indicated in \citealt{2005ATel..599....1K}) as the counterpart of the hard X-ray source \tre.  The magnitudes from 2MASS are $J=12.95\pm0.03$, $H=10.825\pm0.02$, $K=9.80\pm0.02$.

\begin{figure}
\includegraphics[width=8.7cm,angle=0]{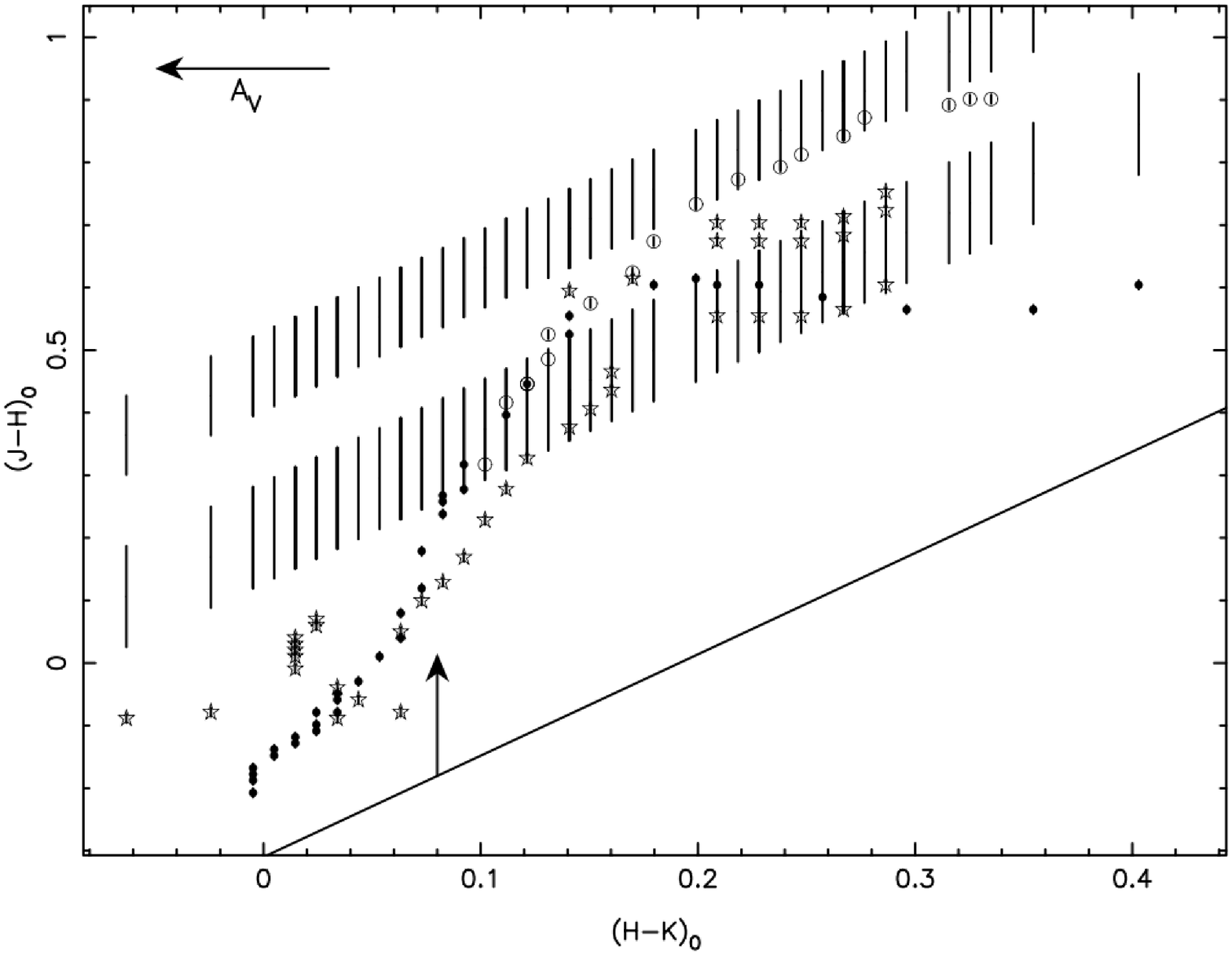}
\caption{  Possible $(J-H)_0$, $(H-K)_0$ combinations allowed by the 2MASS $J, H$ and $K$ magnitudes of the sources \tre (top hatched area), \sei (lower limit to $(J-H)_0$ indicated by the solid line and arrow) and  \otto(bottom hatched area), for different values of the absorption ($A_V$ increasing from right to left). For comparison, the symbols indicate the couples $(J-H)_0$, $(H-K)_0$ for stars of different spectral type (O9 to M7) and luminosity class (I,III and V) (from \citealt{2000asqu.book..143T} ): dots represent main sequence stars, empty circles are red giants and stars are supergiants. }
\label{NIRcol}
\end{figure}

Figure \ref{NIRcol} shows a comparison  between the intrinsic NIR colours $(J-H)_0$ and $(H-K)_0$ of the counterpart to \tre  and the same colours for typical stars of luminosity class I, III and V and spectral type from O9 to M7 (from \citealt{2000asqu.book..143T}).  The intrinsic NIR colours for the counterpart are obtained  from the 2MASS magnitudes for different values of $A_V$. The comparison is constructed as follows: we assume $(H-K)_0$ as for the tabulated spectral types and calculate the $A_V$ that is required to obtain such an intrinsic colour from the observed $(H-K)$ ($A_V$ related to $A_J$, $A_H$ and $A_K$ as for \uno, with the typical central wavelength of 2MASS filters from \citealt{2006AJ....131.1163S}).  With this  $A_V$, $(J-H)_0$ is derived from the observed $(J-H)$. We accounted for the difference in the photometric system employed by \citet{2000asqu.book..143T} and the 2MASS {\it J,H,K} \footnote{following the transformations at http://www.astro.caltech.edu/~jmc/2mass/v3/transformations/}. 
Interestingly, the  possible combinations of $(J-H)_0$ and $(H-K)_0$ obtained for \tre seem not to agree with the spectral classification as a O9.5 Iab. The NIR colours point instead toward a late type red giant,  or a spectral type not included in the comparison such as a supergiant earlier than O9. The comparison method has been tested by obtaining the NIR colours for objects with a known spectral type: we tested all the sources classified in  \citet{2008A&A...486..911N} (IGR J16465-4507, AX J1841.0-0536, 4U 1907+09, IGR J19140+0951), with  IGRJ 17544-2619 and  XTEJ17391-3021 \citep{2006ESASP.604..165N}, IGRJ16207-5129 \citep{2007A&A...461..631N}, HD 306414 \citep{2005ATel..470....1N} and with the sources \uno and \undici included in this paper (see section \ref{uno} and \ref{undici}).   The agreement is good for all the systems but the O8 Ia type XTEJ17391-3021, which is offset by $\sim0.1~\rmn{mag}$ from to the closest spectral type in our reference table, an O9 I object. The test source 4U 1907+09 has the same spectral type (O9.5 Iab) as \tre and indeed its NIR colours are fully consistent with the spectral classification, while those of \tre are not.
This discrepancy is difficult to explain. There is no indication that the 2MASS photometry is subject to additional uncertainties and  it seems unlikely that the spectra are compatible with that of a late-type giant. We conclude that the counterpart to \tre is peculiar in the NIR region of the spectrum.

\subsection{IGR J16500-3307: an Intermediate Polar}

\quattro was discovered by \integral \citep{2006ApJ...636..765B} and has been associated with the {\it ROSAT} bright object 1~RXS~J164955-330713 \citep{1999A&A...349..389V}. It has been also observed by \swift \citep{2008A&A...482..113M}.

The USNO~A2-0  object U0525-24170526 has been proposed as a possible optical counterpart to  \quattro based on the X-ray position from \integral and {\it Swift/XRT}. An optical spectrum of this source is presented in \citet{2008A&A...482..113M} and is compatible with \quattro being an intermediate polar (IP) CV.  The source is included in the study of hard X-ray detected magnetic CVs by \citet{2010MNRAS.401.2207S}.

We observed the source with \chan for $\sim1.2~\rmn{ks}$ on 2007~Sept.~29, detecting a single source (198 counts) compatible with the \integral, \swift and {\it ROSAT} positions, at \ra=~$16^{\rmn{h}}$~$49^{\rmn{m}}$~$55\fs7$, \dec=~$-33\degr$~$07\arcmin$~$02\farcs 3$.

\begin{figure*}
\begin{minipage}[b]{8.5 cm} 
\centering
\includegraphics [width=5.6 cm]{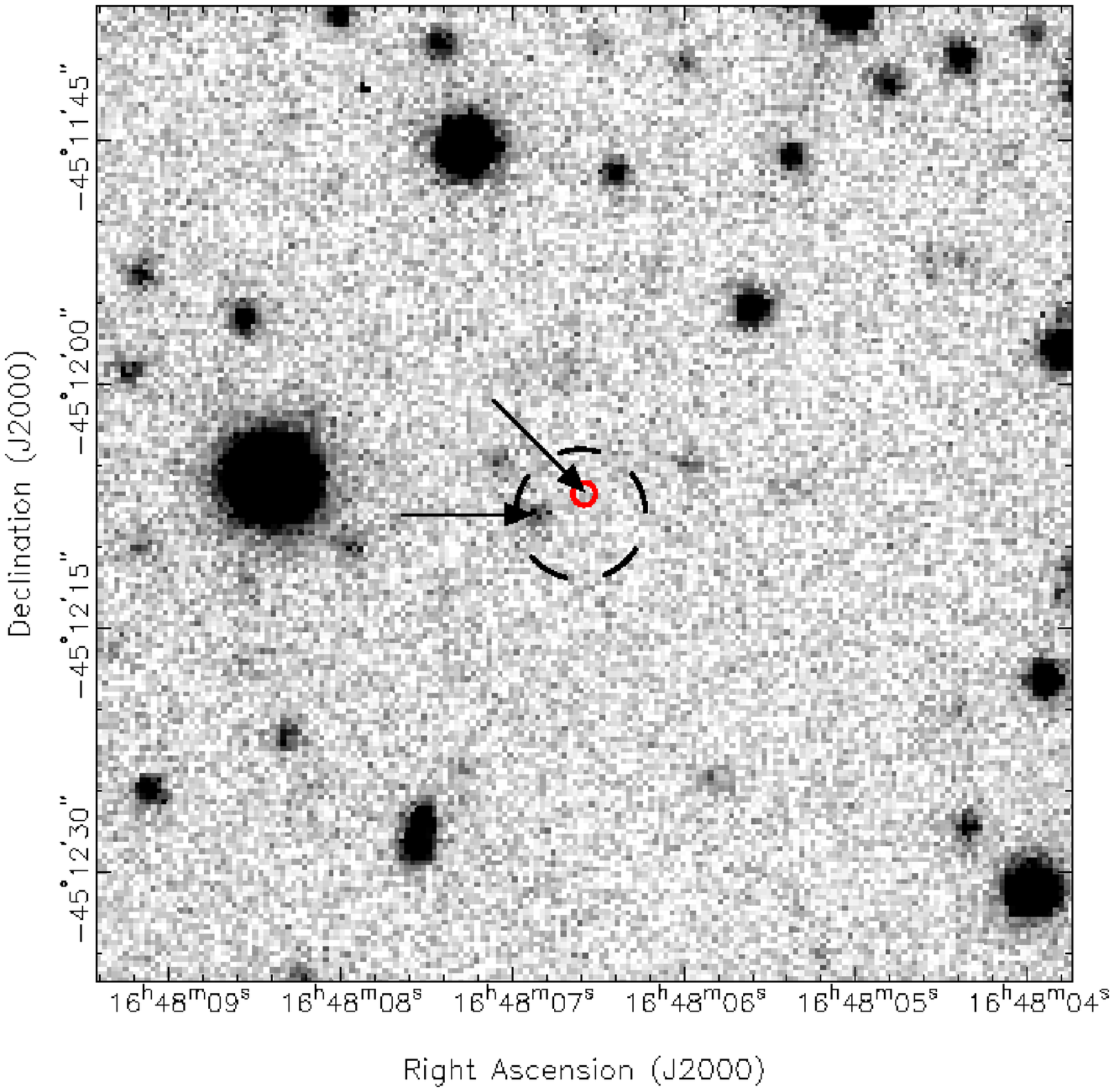}
\caption{\tre, {\it MOSAIC~II}, 300 $\rmn{s}$ in the {\it i'} band. Red error circle: \chan position. Dashed error circle: \xmm. The arrows indicate the position of the candidate counterparts 1 and 2 in \citet{2008A&A...484..783C}.\newline}
\label{fifth}
\end{minipage}
\hspace{0.5 cm}
\begin{minipage}[b]{8.5 cm}         
\centering
\includegraphics[width=5.8 cm]{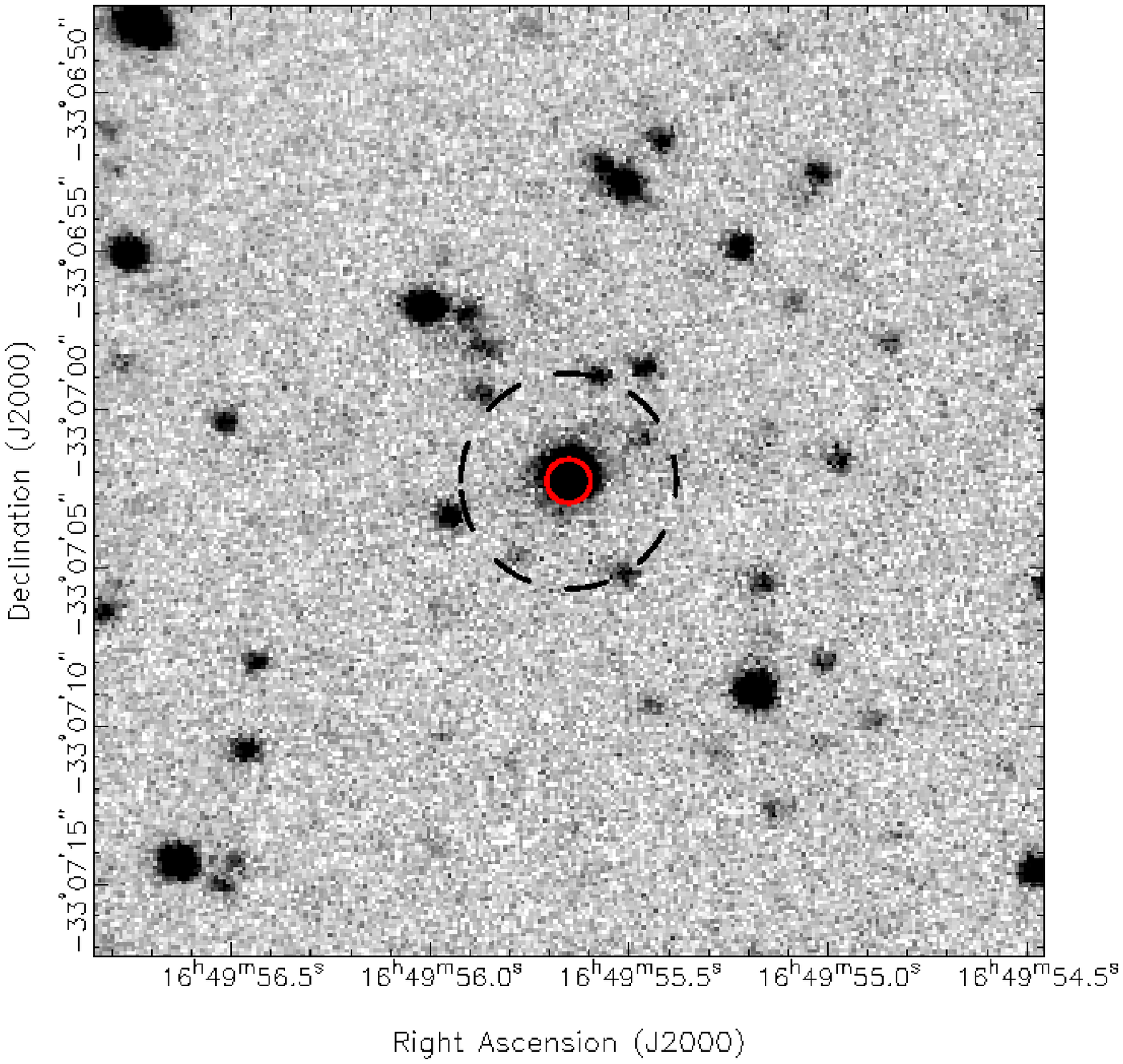}
\caption{\quattro, {\it PANIC}, 75 $\rmn{s}$ in the $K_s$ band. Red error circle: \chan position. Dashed error circle: \swift. \newline \newline \newline}
\label{sixth}
\end{minipage}
\end{figure*}

 Follow-up observations, performed with {\it Blanco/MOSAIC~II} in the {\it r'} band on 2006~Jun.~24 and with {\it Magellan/PANIC} in the $K_s$ band on 2006~Aug.~3, show a bright source inside the \chan error circle (see Figure \ref{sixth}), at \ra=~$16^{\rmn{h}}$~$49^{\rmn{m}}$~$55\fs633$, \dec=~$-33\degr$~$07\arcmin$~$02\farcs 13$ ($\pm~0.09$ arcsec on both coordinates). This position corresponds to the previously proposed counterpart from \citet{2008A&A...482..113M}. The star is also reported in  2MASS as 16495564-3307020, with magnitude $J=14.409~\pm~0.039$, $H=13.969~\pm~0.044$ and $K=13.712~\pm0.049$. After absolute photometric calibration, we measure a magnitude $K_{s}=~13.64~\pm~0.04$ with {\it PANIC} and $r'=~15.94~\pm0.04$ with {\it MOSAIC~II}. The probability that a star with that brightness falls by chance in the \chan error circle is very low ($\sim3\times10^{-7}$ in {\it PANIC} observations). Based on its position, we confirm the object USNO~A2-0~U0525-24170526/2MASS~16495564-3307020  (first proposed by \citealt{2008A&A...482..113M})  as the counterpart of the X-ray source IGR J16500-3307.

\subsection{XTE 1710-281: an eclipsing LMXB}
\label{cinque}

\cinque was serendipitously discovered in 1998 by {\it RXTE/PCA} and associated with the {\it ROSAT} source 1RXS~J171012.3-280754 \citep{1998IAUC.6998....2M}. The source was detected by {\it INTEGRAL/IBIS}  \citep{2004AstL...30..382R} and recently by \xmm (\citealt{2008yCat..34930339W}; \citealt{2009A&A...502..905Y}).
Complete X-ray eclipses and dips have been detected in the {\it RXTE/PCA} light curves, indicating an orbital period of $\sim3.28$ hours. Thermonuclear type I X-ray bursts indicate the object is a NS and strongly suggest that the system is an LMXB \citep{1995xrbi.nasa..175L}.  The distance $d$ has been constrained from type I X-ray bursts: \citet{2001AAS...199.2704M} indicate $d=15-20~\rmn{kpc}$, while \citet{2008ApJS..179..360G} obtain $d=12-16~\rmn{kpc}$. 

 \cinque is reported in the second \xmm serendipitous source catalog \citep{2008yCat..34930339W} at \ra=~$17^{\rmn{h}}$~$10^{\rmn{m}}$~$12\fs532$, \dec=~$-28\degr$~$07\arcmin$~$50\farcs 95$,  
with an accuracy of 1 arcsec at $1\sigma$ on both coordinates. 
Taking advantage of this recent position we performed a search for the counterpart in {\it I} band, observing  on 2006~Aug.~3 with {\it Magellan/IMACS}. We detect one object inside the 90 per cent confidence radius around the \xmm position (see Figure \ref{seven}) at \ra=~$17^{\rmn{h}}$~$10^{\rmn{m}}$~$12\fs6$, \dec=~$-28\degr$~$07\arcmin$~$51\farcs 0$ (~$\pm~0.1$ arcsec on both coordinates). Its magnitude in {\it I} band is $I=19.7~\pm~0.1$. The probability that this source falls by chance inside the \xmm error circle is $2.3\times10^{-4}$.
Since the distance of the source has been constrained, we can infer an upper and lower limit to the absolute magnitude $M_{I}$ of that candidate counterpart in {\it I} band.
 We derive the absolute extinction coefficient in the {\it I} band similarly to what  we did for \due (see Section \ref{due}), assuming the $N_H$ obtained by  \citet{2009A&A...502..905Y} from \xmm spectra.
Considering a  distance $12~\rmn{kpc}<d<20~\rmn{kpc}$  (see above), the {\it I} band absolute magnitude of the counterpart to \cinque is in between $M_{I}\sim3.43$ and $M_{I}\sim2.32$. This is in agreement with what is expected if we are observing the disk of an high inclination LMXB \citep{1995xrbi.nasa...58V} and supports the association  with XTE 1710-281. 

\subsection{XTE J1716-389: an obscured HMXB system }
\label{sei}
  \sei was discovered by {\it RXTE} between 1996 and 1997 \citep{1999MmSAI..70..881R} and corresponds to the source KS1716-389 \citep{2006MNRAS.366..918C} detected two years before the launch of {\it RXTE} itself by the {\it TMM/COMIS} telescope on board the {\it Mir-Kvant} module \citep{1995AstL...21..431A} . It is also reported in the {\it EXOSAT} Slew survey catalogue \citep{1999A&AS..134..287R} as EXO~J1715557.7-385 and is associated with the {\it ROSAT} source 1RXH~J171556.7-385150. It has been observed with {\it ASCA} and detected in hard X-rays by \integral \citep{2006ApJ...636..765B}.

 Extensively monitored by  {\it RXTE}, the system has shown a highly-variable persistent emission (\citealt{1999AAS...195.3912W};  \citealt{2006MNRAS.366..918C}). It presents dips with a duration of $\sim30$ days and a recurrence period of $\sim100$ days, associated with sudden increases in the absorption column density $N_{H}$. Even outside the dipping phase the $N_{H}$ is high ($\sim 10^{23}~\rmn{cm}^{-2}$) compared to the Galactic value towards the source ($\sim2\times10^{22}~\rmn{cm}^{-2}$), indicating that the system is absorbed locally. The source presents remarkable similarities with the class of obscured HMXBs (\citealt{1999AAS...195.3912W}; \citealt{2006A&A...453..133W}). The $\sim100$-day recurrence of the dips is likely not associated with the system orbital period, but with a super-orbital periodicity \citep{1999AAS...195.3912W} as has been observed in many HMXBs with a supergiant companion.

\citet{2005A&A...432L..49S} present a search for an optical/NIR counterpart to \sei based on its {\it ROSAT} position. They indicate a candidate counterpart in optical, also reported in 2MASS and a few NIR sources inside the {\it ROSAT/HRI} error circle.

We observed \sei with {\it Chandra/ACIS} for $\sim1.1~\rmn{ks}$ on 2008~Sep.~23 detecting one faint source (12 counts in a $0.3-7~\rmn{keV}$ energy band) compatible with the {\it ROSAT} pointing from \citet{2005A&A...432L..49S}, at coordinates \ra=~$17^{\rmn{h}}$~$15^{\rmn{m}}$~$56\fs42$, \dec=~$-38\degr$~$51\arcmin$~$54\farcs 127$. This position excludes the  optical counterpart proposed in \citet{2005A&A...432L..49S}, being $\sim4.5$ arcsec (more than $10\sigma$) far away from it.

\begin{figure*}
\begin{minipage}[b]{8.5 cm} 
\centering
\includegraphics[width=5.9 cm]{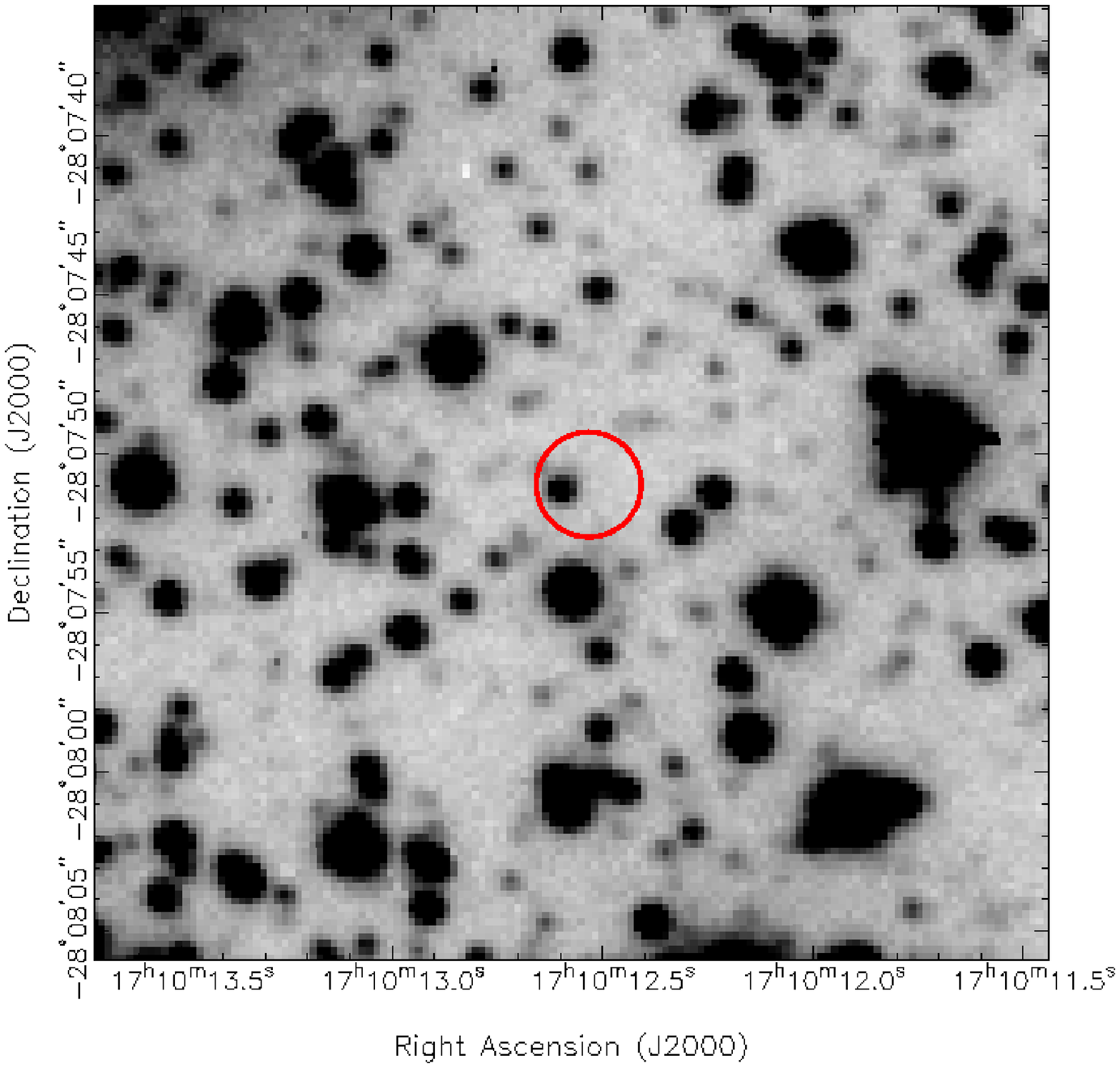}
\caption{\cinque, {\it IMACS}, 300 $\rmn{s}$ in the {\it I} band. Red error circle: \xmm position.\newline \newline \newline }
\label{seven}
\end{minipage}
\hspace{0.5 cm}
\begin{minipage}[b]{8.5 cm}         
\centering
\includegraphics[width=5.4 cm]{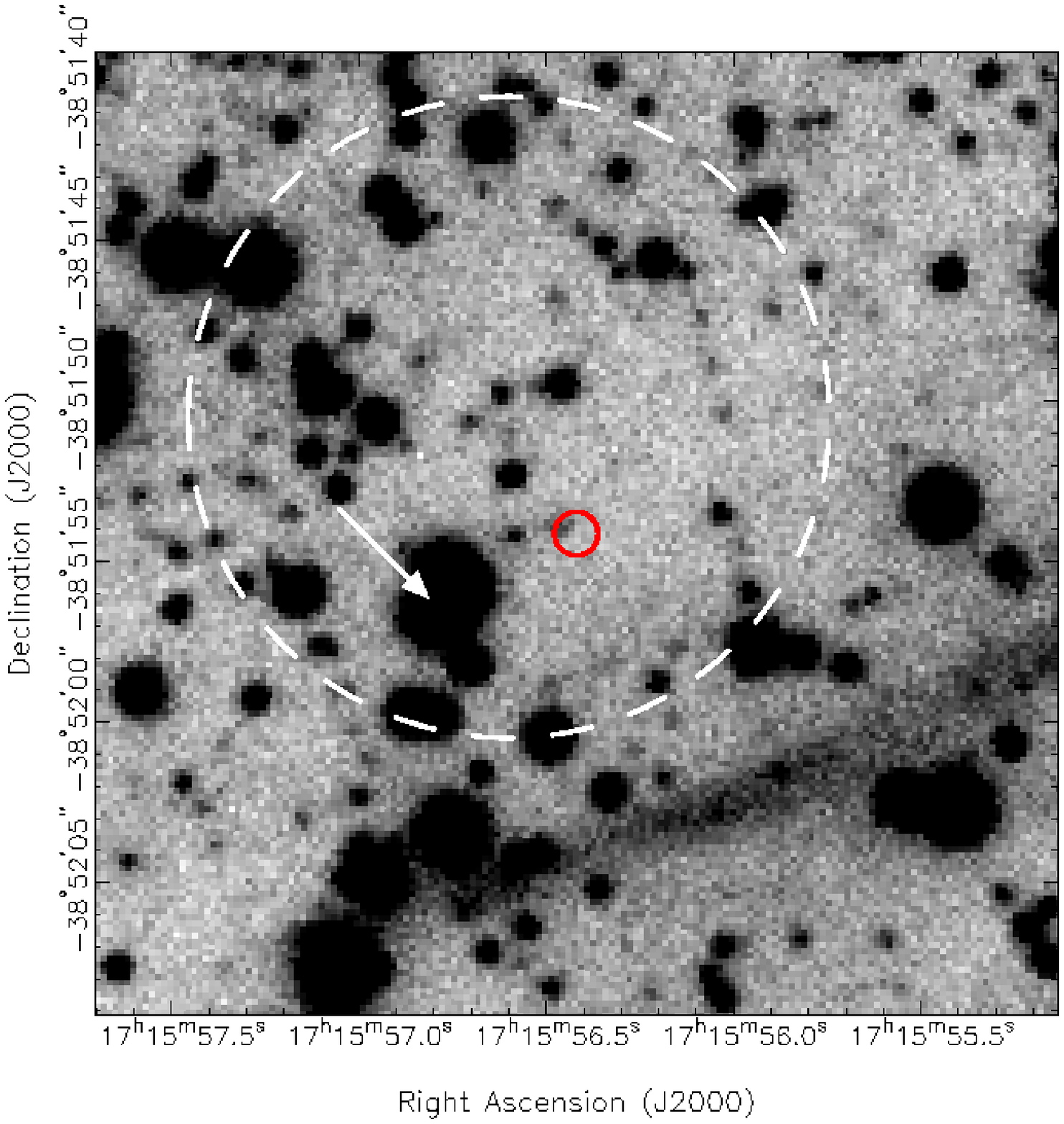}
\caption{\sei, {\it LDSS3}, 300 $\rmn{s}$ in the {\it i'} band. Red error circle: \chan position. Dashed error circle: {\it ROSAT}. The arrow indicates the position of the candidate counterpart proposed in \citet{2005A&A...432L..49S}, which the \chan position rules out. There is one candidate counterpart in the \chan error circle.}
\label{eight}
\end{minipage}
\end{figure*}

 Follow-up optical observations in {\it i'} band were performed on 2009~May~7 with {\it Magellan/LDSS3}, showing a very faint source inside the 90 per cent confidence \chan error circle, at \ra=~$17^{\rmn{h}}$~$15^{\rmn{m}}$~$56\fs457$, \dec=~$-38\degr$~$51\arcmin$~$53\farcs 9$~$\pm 0.1$ arcsec on both coordinates (see Figure \ref{eight}). The probability that the source falls by chance in the \chan error-circle is $\sim3\times10^{-4}$. The object is reported in the 2MASS catalogue, with magnitude $H=13.569~\pm~0.09$, $K=12.579~\pm~0.059$. A limit to the magnitude in {\it J} band is also reported, the object being fainter than $J=15.058$.  We observed the source in a non photometric night: we obtain an estimate of the {\it I}-band magnitude $I\sim22.7$ by calibrating our {\it LDSS3} observation with the zero-point from a previous observing run (see Section \ref{optical}). 
 As for \tre  (see last paragraph in Section \ref{tre}), the possible combinations of $(J-H)_0$ and $(H-K)_0$ allowed by the 2MASS observed colours $(J-H)$ (lower limit) and $(H-K)$ are shown in Figure \ref{NIRcol}. The colours are compatible with a star of any class, but only for an $A_V$ ranging between $\sim8$ and $\sim17$. Those are acceptable values for a HMXB, where the companion star can be highly obscured in the optical due to its own wind ( $A_V\sim9-15$, i.e., for  the five systems investigated in \citealt{2010A&A...510A..61T}). On the other hand, a main sequence star or a red giant are unlikely since the high $A_V$ requires a dense stellar wind to be justified. An even higher $A_V$ ($\sim45$) corresponds to the $N_H$ obtained from the X-rays observations (see above). This indicates that there is absorption in the surroundings of the compact object, not affecting the companion star. We conclude that the source 2MASS 17155645-3851537 is most likely a massive self-absorbed star: this supports its positional association with \sei and the classification of the latter as an obscured HMXB.

\subsection{IGR J17254-3257: a candidate UCXB}

\sette was discovered by \integral  in 2003 \citep{2004ATel..229....1W} and reported in various catalogues of {\it INTEGRAL/IBIS} sources (i.e. \citealt{2004ApJ...607L..33B}).   It is also a {\it ROSAT} source (1RXS~J172525.5-325717), it  has been continuously detected by {\it RXTE/PCA} at very low count-rate since 1999 \citep{2005A&A...432L..49S} and was also observed with \xmm \citep{2007A&A...469L..27C} and \swift \citep{2009AIPC.1126..104C}.

A type I X-ray burst  detected on 2004~Feb.~17 \citep{2006ATel..778....1B} indicated that the system is an LMXB hosting a NS. Moreover, a long thermonuclear burst lasting about 15 minutes was observed on 2006 October 1, placing \sette in the small group of XRBs showing bursts very different in duration (\citealt{2007A&A...469L..27C} and references therein). Based on the persistent behaviour of the source at a low accretion rate, \citet{2007A&A...465..953I} proposed \sette as a candidate UCXB. An upper limit to the distance of 14.5 $\rmn{kpc}$ has been estimated from the bursts \citep{2007A&A...469L..27C}.

 \citet{2009ATel.2032....1Z} reported two possible optical counterparts to \sette that are compatible with the \xmm position.

We observed the field of IGRJ17254-3257 with {\it Magellan/PANIC} on 2006~Aug.~03 in $K_s$ band:  in addiction to the two sources indicated by \citet{2009ATel.2032....1Z} (see Figure \ref{ten}), 11 further object are resolved by our PSF photometry within the \xmm error circle.

Considering a maximum distance to the source of 14.5 $\rmn{kpc}$ and with the $N_H$  from \citet{2007A&A...469L..27C}, we can set a limit to the absolute magnitude $M_K$ of the candidates, similarly to what we did for \cinque (see Section \ref{cinque}). In order to reduce the number of possible counterparts we compared those magnitudes with that of the UCXB 4U~0614+09, for which the case for an ultracompact nature is strong (\citealt{2004MNRAS.348L...7N}; \citealt{2008PASP..120..848S}). For  4U~0614+09,  $d=3.2$ $\rmn{kpc}$ \citep{2009arXiv0909.3391K}, $K=17.1$ and $A_V=1.41$ \citep{2007MNRAS.379.1108R}, thus $M_K=4.46$. This is consistent  with what expected for an UCXB \citep{1995xrbi.nasa...58V}. The first candidate from \citet{2009ATel.2032....1Z} should be very close by ($d=0.6~\rmn{kpc}$) in order to have a similar $M_K\sim4$: this is unlikely because \sette would than be the nearest XRB known and its X-ray luminosity during X-ray bursts would be anomalously low. Beside that, unfortunately the comparison does not provide any constraint to further reduce the number of candidates in the \xmm error circle.
A localization of the X-ray source with \chan is necessary to identify the actual counterpart.

\subsection{XTE J1743-363: a candidate SFXT}

\otto was discovered with \rxte in 1999 \citep{1999IAUC.7120....1M}. The system has been detected by {\it INTEGRAL/IBIS} several times in 2004 at diverse flux levels (\citealt{2004AstL...30..382R}; \citealt{2004ATel..332....1G}). It also showed a few-hour long outburst, because of which \otto is considered a candidate SFXT \citep{2006ApJ...646..452S}. No search for an optical or NIR counterpart is reported in the literature.

We observed \otto with {\it Chandra/HRC} on 2008~Feb.~8 for $\sim1.2~\rmn{ks}$ detecting one faint source (11 counts) compatible with the \integral position (from \citealt{2006ApJ...636..765B}) at coordinates \ra=~$17^{\rmn{h}}$~$43^{\rmn{m}}$~$01\fs3$, \dec=~$-36\degr$~$22\arcmin$~$22\farcs 0$.

\begin{figure*}
\begin{minipage}[b]{8.5 cm} 
\centering
\includegraphics [width=5.7 cm]{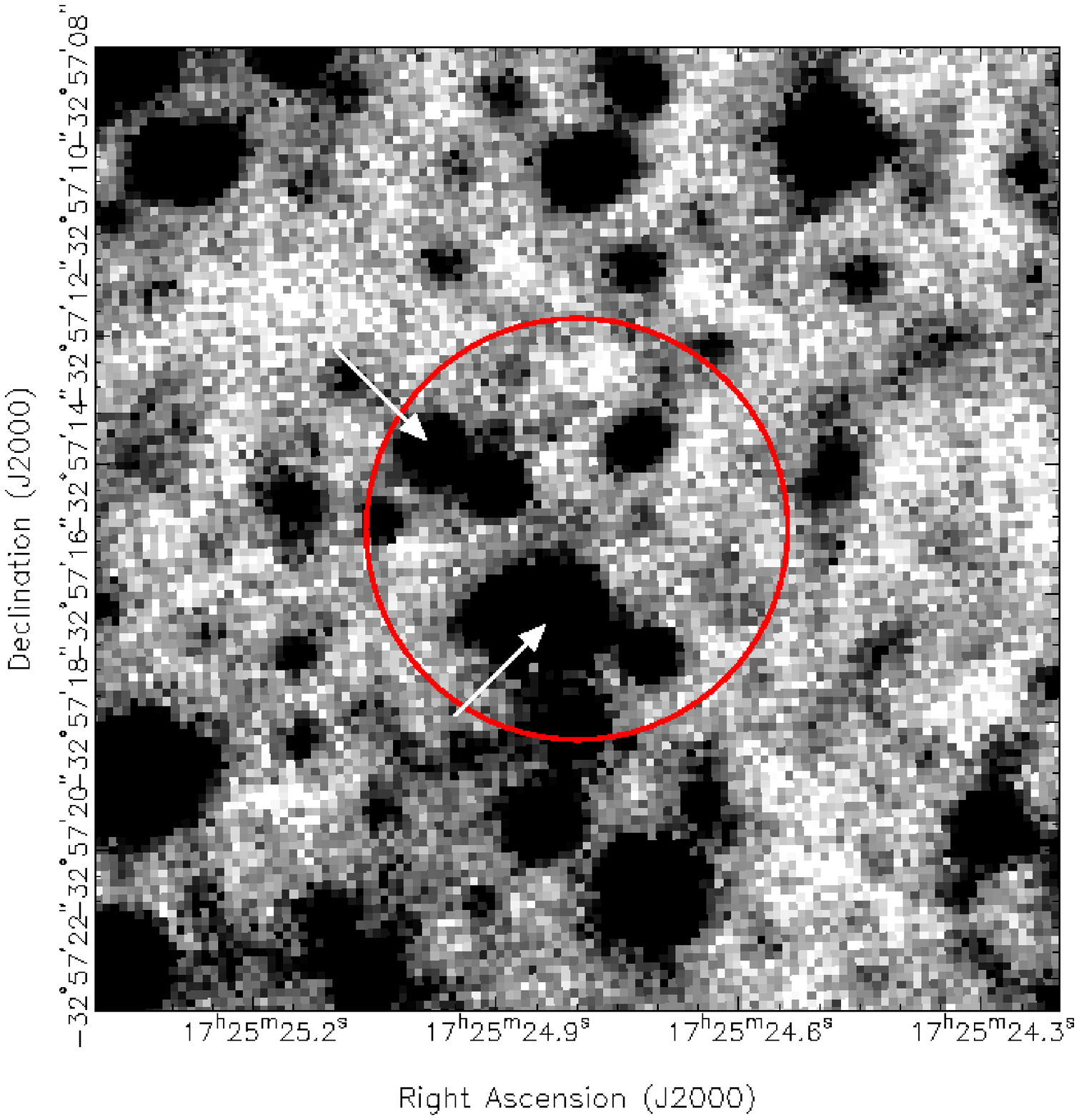}
\caption{ \sette: {\it PANIC}, 15 $\rmn{s}$ in the $K_s$ band. Red error circle: \xmm position. The arrows indicate the position of the two candidate counterparts in  \citet{2009ATel.2032....1Z} (first and second from bottom to top).}
\label{ten}
\end{minipage}
\hspace{0.5 cm}
\begin{minipage}[b]{8.5 cm}         
\centering
\includegraphics[width=5.8 cm]{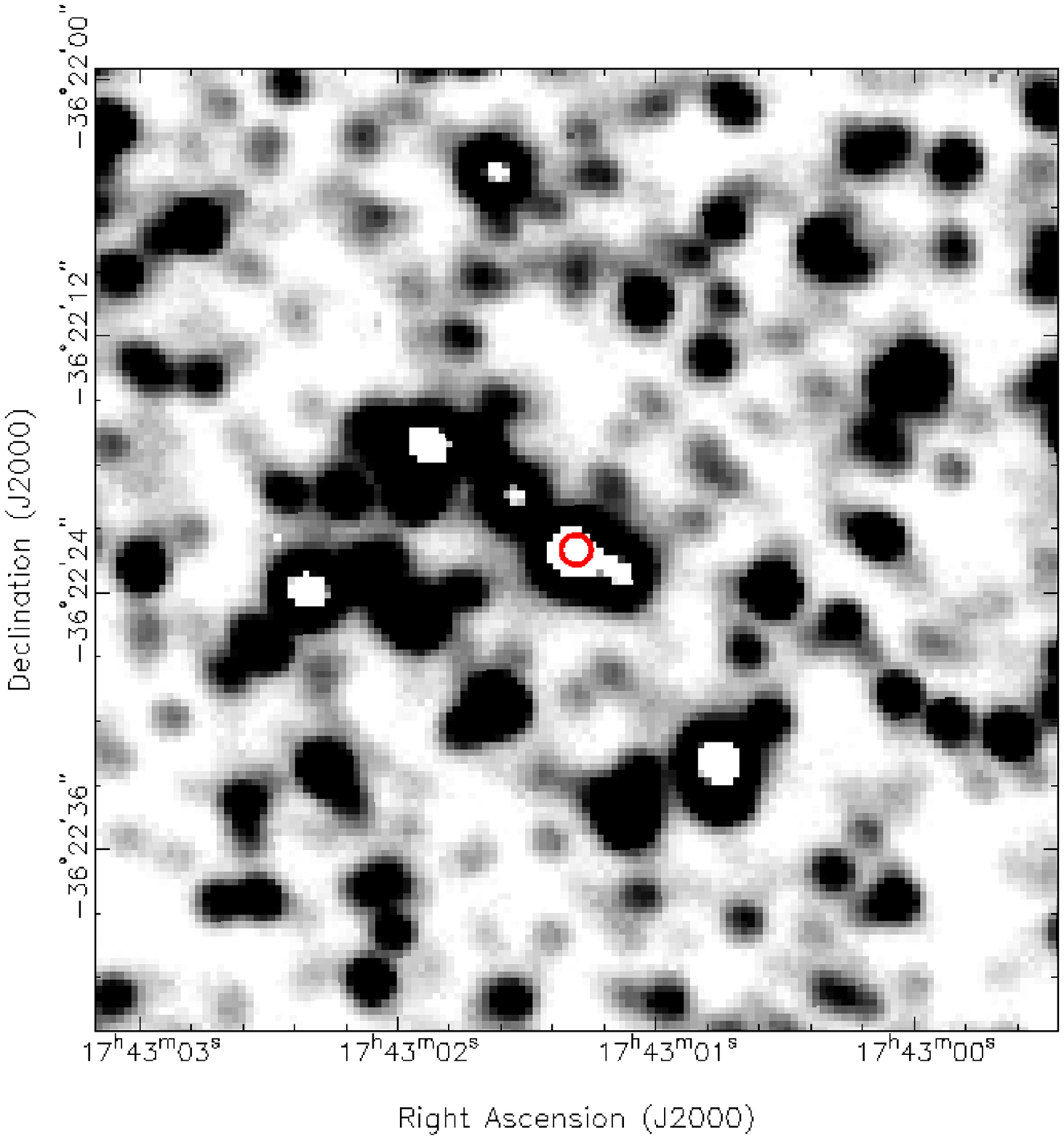}
\caption{\otto: {\it EMMI}, 600 $\rmn{s}$ in the {\it I} band. Red error circle: \chan position.\newline \newline}
\label{eleven}
\end{minipage}
\end{figure*}

 Optical images in the {\it I} band, collected with {\it EMMI} on 2007~June~22, show a bright star lying inside the \chan error circle (see Figure \ref{eleven}) at \ra=~$17^{\rmn{h}}$~$43^{\rmn{m}}$~$01\fs324$, \dec=~$-36\degr$~$22\arcmin$~$22\farcs 2$ ($\pm 0.1$ arcsec on both coordinates). The position of the source is coincident within the error with the object 2MASS~17430133-3622221, for which the following magnitudes are reported in the catalogue: $J=9.616~\pm~0.024$, $H=8.305~\pm~0.034$, $K=7.624~\pm~0.026$.   We cannot report a magnitude in {\it I} band since the field was observed in a non-photometric night. 

Given the classification as a SFXT, the companion star in \otto is expected to be a supergiant, locally absorbed in the NIR due to its own wind (see Section \ref{intro} and \ref{sei}). As for \tre and \sei, figure \ref{NIRcol} shows the combination of $(J-H)_0$ and $(H-K)_0$ allowed by the observed 2MASS colours, for different values of the absorption. The colours are compatible with a type G0-6 III  ($A_V$ beween 8 and 9) or G/K I  ($A_V$ between 6 and 9)  and do not exclude a supergiant of spectral type earlier than O9 (see last paragraph in section \ref{tre}) if $A_V>12$. As in the case of \sei, a main sequence or giant star are allowed by the colours, but unlikely due to the high $A_V$.

Based on its position we conclude that the object 2MASS~17155645-3851537 is most likely the NIR counterpart to \sei. The NIR colours of the counterpart are consistent with  a late type supergiant and do not exclude an early O I type. This is consistent with the classification of the X-ray source as a SFXT and supports the counterpart association.

\subsection{IGR J17597-220: a dipping LMXB}

\nove was first detected in 2001 by {\it RXTE/PCA} \citep{2003ATel..156....1M} but it was reported for the first time in 2003  \citep{2003ATel..155....1L} as a new \integral source. For that reason it is usually indicated as either \nove or XTE J1759-220.
Type I X-ray bursts from the source have been observed by {\it INTEGRAL/JEM-X}, identifying the compact object as a NS \citep{2007ATel.1054....1B} and the system as a probable LMXB. 
\nove has also shown dips of $\sim$30 per cent with a duration of $\sim5$ minutes, from which \citet{2003ATel..156....1M} suggested an orbital period of 1-3 $\rmn{hours}$.

\xmm observations localised the source with a 4 arcsec accuracy \citep{2006A&A...453..133W}. \citet{2008A&A...484..783C} identified 6 candidate counterparts consistent with the \xmm position on NIR observations in {\it J}, {\it H} and $K_{s}$ bands. We detect a single {\it Chandra/HRC} source (227 counts) inside the \xmm error circle, at \ra=~$17^{\rmn{h}}$~$59^{\rmn{m}}$~$45\fs52$, \dec=~$-22\degr$~$01\arcmin$~$39\farcs 17$, during a $\sim1.2~\rmn{ks}$-long observation performed on 2007~Oct.~23. 

Follow-up observations in {\it I} band  were performed with {\it NTT/EMMI} on 2006~June~22 and with {\it Magellan/LDSS3} in {\it i'} band on 2009~May~7. A single, faint source lies inside the \chan error circle in both bands, at \ra=~$17^{\rmn{h}}$~$59^{\rmn{m}}$~$45\fs525$, \dec=~$-22\degr$~$01\arcmin$~$39\farcs 25$ ($\pm 0.1$ arcsec on both coordinates). The detection is evident in the 300 $\rmn{s}$-long {\it LDSS3} images (see Figure \ref{thirteen}), while it is less significant in the 600 $\rmn{s}$-long {\it EMMI} one. We consider the detection with {\it EMMI} as real due to its positional coincidence with that of the source in {\it LDSS3}.  The observing nights with both instruments were not photometric: we obtain an estimate of the {\it I}-band magnitude $I\sim22.4$ by calibrating our {\it LDSS3} observation with the zero-point from a previous observing run (see Section \ref{optical}).
The probability that our candidate counterpart falls by chance inside the \chan error circle is $\sim 1\times10^{-4}$ and $\sim 8\times10^{-4}$ for {\it LDSS3} and {\it EMMI} respectively (photometry on a smaller field for {\it EMMI}). Its position matches that of the Candidate $1$ in \citet{2008A&A...484..783C} that we establish as the very likely optical/NIR counterpart of \nove. 

\subsection{IGR J18490-0000: a Pulsar Wind Nebula}

 A Pulsar Wind Nebula (PWN) is a nebula powered by the interaction of the highly relativistic particle wind formed in the magnetosphere of a pulsar with the surrounding material. 
\dieci was discovered by \integral in the spring of 2003, during a survey of the Sagittarius arm tangent region of the Galaxy \citep{2004AstL...30..534M}.  In the soft X-rays the source is composed of a point-like source surrounded by an extended nebula \citep{2008AIPC.1085..312T}. Its morphology and spectral properties at X-rays are reminiscent of a PWN, although pulsations have not been detected so far \citep{2009AIPC.1126..259M}. The association of \dieci with a PWN was further strenghtened  by the discovery of a $\rmn{TeV}$ counterpart with the High Energy Stereoscopic System {\it HESS} \citep{2008AIPC.1085..312T}.

{\it Swift/XRT} observations are presented in \citet{2008A&A...482..731R}: based on the \swift position, the object 2MASS~18490182-0001190 has been proposed as a possible NIR counterpart \citep{2008A&A...482..731R}. 

 A $\sim$1.2 $\rmn{ks}$-long {\it Chandra/HRC} observation of the field, obtained on 2008~Feb.~16, shows a single source (22 counts) inside the \swift error circle, at \ra=~$18^{\rmn{h}}$~$49^{\rmn{m}}$~$01\fs59$, \dec=~$-00\degr$~$01\arcmin$~$17\farcs 73$, whose morphology is compatible with an extended nebula. Those coordinates exclude the association of \dieci with the candidate counterpart proposed by \citet{2008A&A...482..731R}, which is located at $\sim$3.8 arcsec (over 9$\sigma$) from the \chan position.

\begin{figure*}
\begin{minipage}[b]{8.5 cm} 
\centering
\includegraphics[width=5.9 cm]{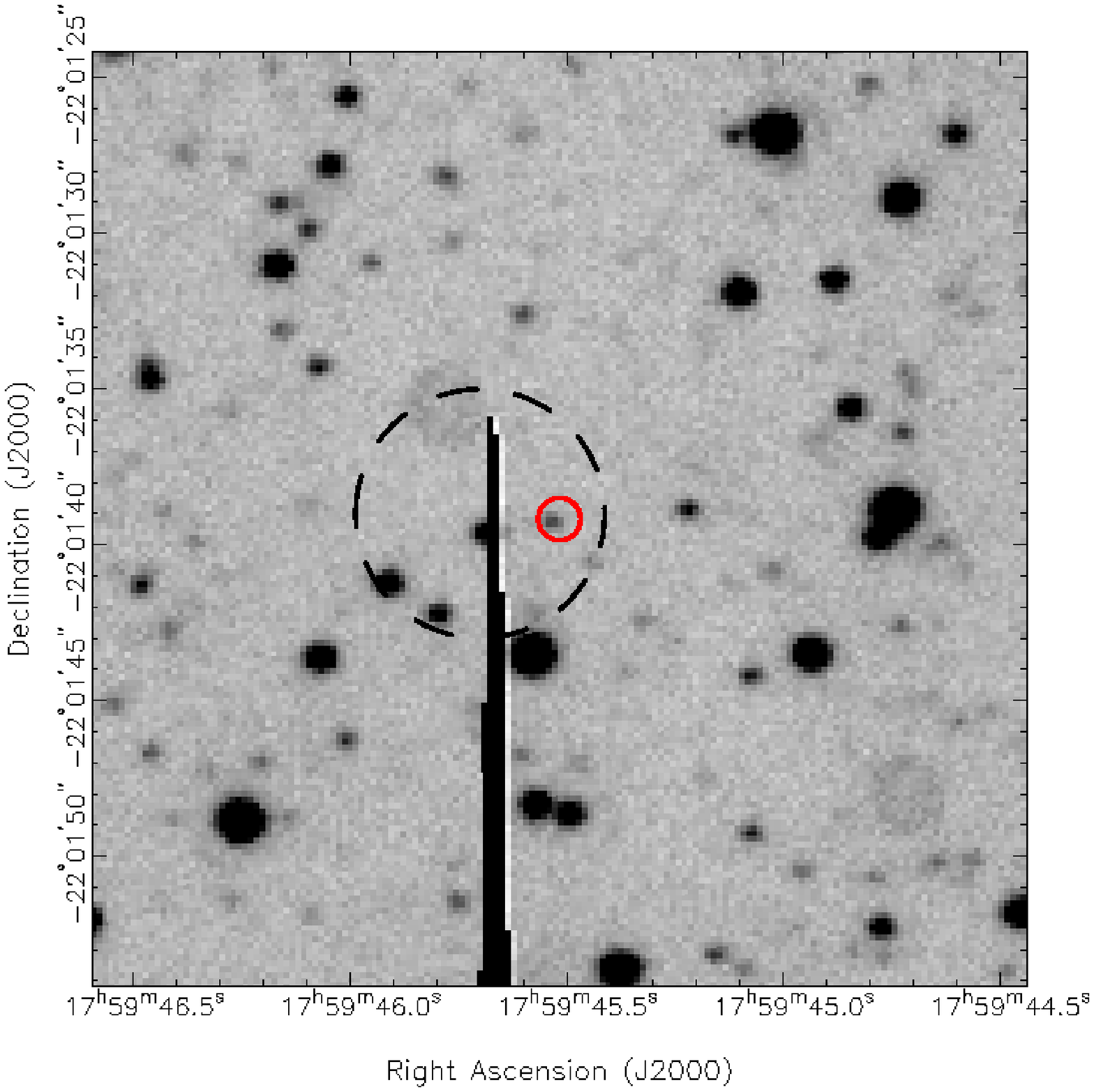}
\caption{ \nove: {\it LDSS3}, 300 $\rmn{s}$ in the {\it i'} band. Red error circle: \chan position. Dashed error circle: \xmm \newline \newline}
\label{thirteen}
\end{minipage}
\hspace{0.5 cm}
\begin{minipage}[b]{8.5 cm}         
\centering
\includegraphics[width=5.6 cm]{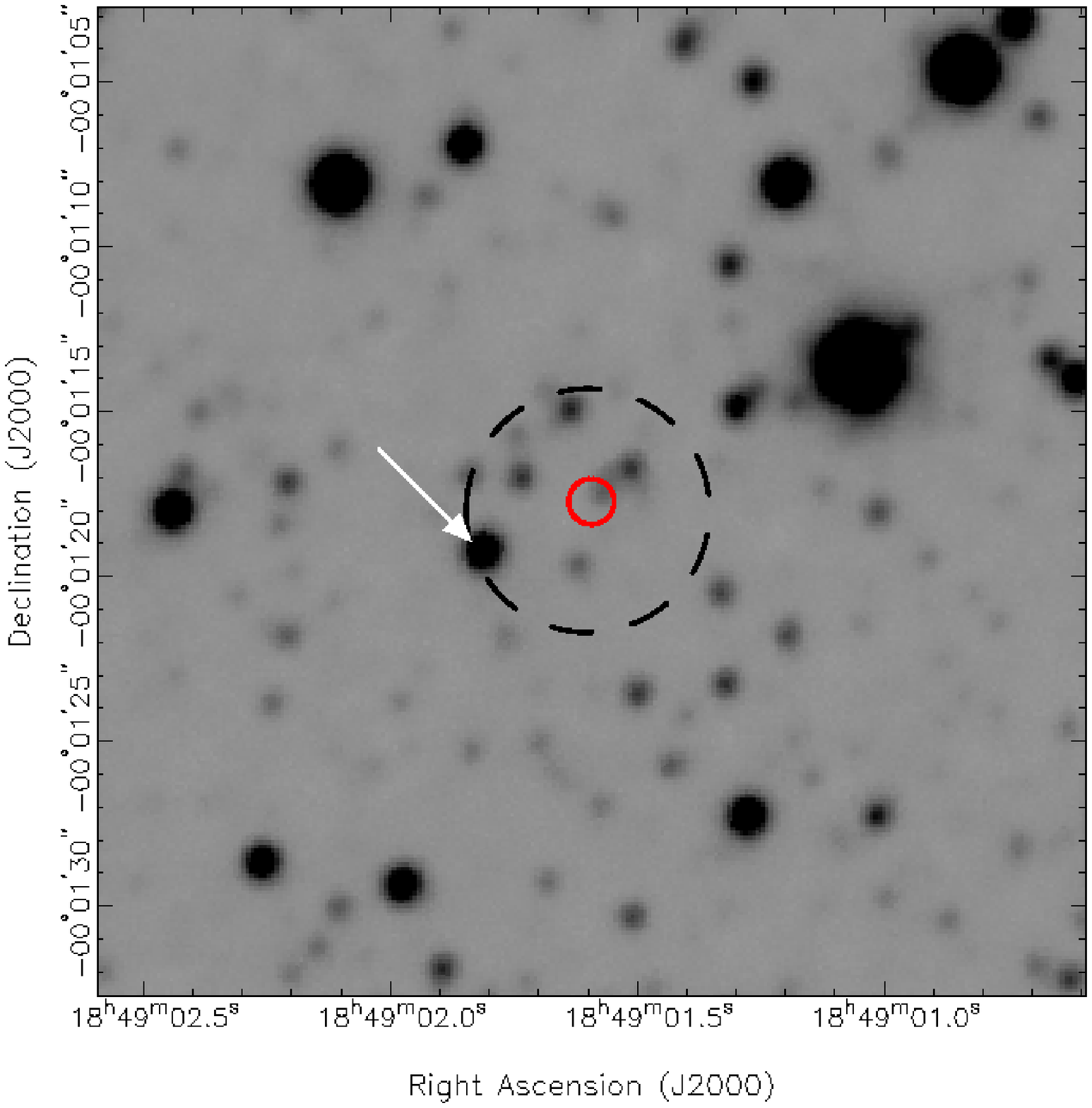}
\caption{\dieci: {\it PANIC}, 15 $\rmn{s}$ in the $K_s$ band. Red error circle: \chan position. Dashed error circle: \swift. The arrow indicates the candidate counterpart from \citet{2008A&A...482..731R}}
\label{sixteen}
\end{minipage}
\end{figure*}

We observed the field of \dieci with {\it Blanco/MOSAIC~II} on 2006~June~25 in {\it i'} band and did not detect any optical counterpart. Nevertheless, giving the low number of sources that we observe in the field and comparing the {\it i'} band images with 2MASS infrared ones, we consider it likely that a dark cloud is located between us and the source, obscuring the counterpart.  Further observations in the $K_{s}$ band,  performed on 2009~Jul.~16 with {\it Magellan/PANIC,} revealed a faint candidate counterpart on the edge of the 90 per cent \chan error circle (Figure \ref{sixteen} ) at \ra=~$18^{\rmn{h}}$~$49^{\rmn{m}}$~$01\fs563$, \dec=~$-00\degr$~$01\arcmin$~$17\farcs 35$ ($\pm~0.1$ arcsec on both coordinates). After absolute photometric calibration, we measure a magnitude of $K_{s}=16.4~\pm~0.1$. 

The object does not look extended as one would expect for a PWN: our PSF fitting indicates a point-like source, which looks partially blended with a nearby star (Figure \ref{sixteen}). The two sources are resolved by the PSF fitting. This suggests that the object is a foreground star, although its positional coincidence with \dieci within the accuracy of \chan has a low probability of being due to chance ($\sim1.6\times10^{-5}$). We encourage spectroscopic observations of the source in order to investigate its association with IGR J18490-0000.

\subsection{IGR J19308+0530: an L/IMXB with an F8 companion}
\label{undici}

\undici was discovered by \integral \citep{2006ApJ...636..765B} and observed by \swift in  X-rays and in the UV band 170-650 $\rmn{nm}$ \citep{2008A&A...482..731R}. 

Based on the \swift position, \citet{2008A&A...482..731R} identify the star TYC~486-295-1/2MASS~J19305075+0530582 as a possible counterpart. This object is classified as an F8 star in the survey by \citet{1949ApJ...109..426M}. Based on the typical parameters of an F8 star, \citet{2008A&A...482..731R} suggest \undici is a L/IMXB in quiescence or a CV at a distance of $\sim1$ $\rmn{kpc}$ or lower. Fitting the \swift spectrum with a black body of temperature $kT=$0.2 $\rmn{keV}$ the authors obtained a 2-10 $\rmn{keV}$ luminosity of $\sim4\times10^{31}$ $\rmn{erg\,s^{-1}}$ at 1 $\rmn{kpc}$. The corresponding luminosity in the 0.5-10 $\rmn{keV}$ range is $\sim4\times10^{33}$ $\rmn{erg\,s^{-1}}$. The spectrum is very soft, suggesting \undici is most likely not a CV \citep{2006ApJ...646L.143P} but an L/IMXB hosting a NS or a BH in quiescence. This suggestion is strengthened by the fact that the spectra of NS/BH LMXBs in quiescence at a 0.5-10 $\rmn{keV}$ luminosity level of $\sim10^{33}$ $\rmn{erg\, s^{-1}}$ are expected to be dominated by the soft black-body component, as found by \citet{2004MNRAS.354..666J} and updated in \citet{2008AIPC..983..519J}.

 In a  $\sim$1.1\,$\rmn{ks}$-long {\it Chandra/HRC} observation on 2007~Jul.~30, we detected a single source (26 counts) inside the \swift error circle, at \ra=~$19^{\rmn{h}}$~$30^{\rmn{m}}$~$50\fs77$, \dec=~$05\degr$~$30\arcmin$~$58\farcs 09$.

\begin{figure*}
\begin{minipage}[b]{8.5 cm} 
\centering
\includegraphics[width=5.8 cm]{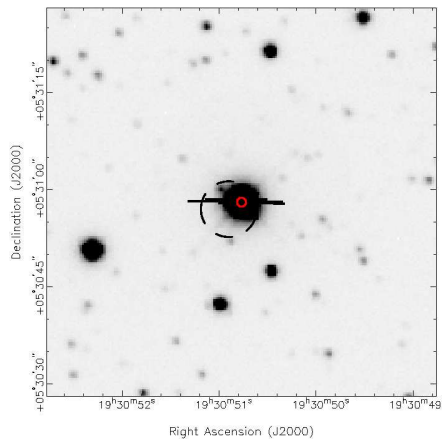}
\caption{ \undici: {\it MOSAIC~II}, 10 $\rmn{s}$ in the {\it I} band. Red error circle: \chan position. Dashed error circle: \swift}
\label{seventeen}
\end{minipage}
\end{figure*}

We searched for an optical counterpart with {\it Blanco/MOSAIC~II} in {\it r} band, on 2006~Jun.~22: the \chan error circle includes a very bright star that is saturated even in the 10 $\rmn{s}$-long image (Figure \ref{seventeen}). Its position is compatible with the position of the previously proposed F8-type counterpart as reported in the Tycho catalogue, in 2MASS, in the LF Survey catalogue and also in UCAC 2 and 3, if the motion of the source since the epoch of each catalogue to that of our observations in 2006 is taken into account. The object has a proper motion of -2.9$~\pm~$0.6 $\rmn{mars\,yr^{-1}}$ in \ra$~$and -10.5$\pm$0.5 $\rmn{mars\,yr^{-1}}$ in \dec$~$(from UCAC3). The magnitudes reported in 2MASS are $J=9.617~\pm~0.032$ (poor photometry) $H=9.245~\pm~0.023$ and $K= 9.130~\pm~0.023$. The intrinsic NIR colours (obtained with the method used for \tre, \sei and \otto) are consistent with the spectral classification. Moreover, the Supplement-1 to the Tycho-2 catalogue reports $B_T=11.706$ and $V_T=10.915$. 
We confirm the association of \undici with the F8 star TYC~486-295-1$/$2MASS~J19305075+0530582  ( first proposed by  \citealt{2008A&A...482..731R}) and we suggest the source is most likely an L/IMXB in quiescence. 

\section{Conclusion}

We have investigated a sample of 11 Galactic X-ray sources recently discovered by \integral or \rxte. For 9 of those, we presented a refined position from \chan observations (Table \ref{chandra}), localising the targets with a positional accuracy of 0.6 arcsec at a 90 per cent confidence level. Thanks to the accurate X-ray position, we have detected a counterpart for all the sources we observed with \chan:  the previously proposed counterparts to \uno, \quattro, \undici and \tre are confirmed by our observations, supporting their classification as, respectively, a SyXB, a CV, an L/IMXB and a SFXT ( altough we evidenced some peculiarity in the NIR colours of the latter). The counterpart to the obscured source \sei is consistent with it being a HMXB. The NIR colours of the counterpart to the SFXT candidate \otto indicate indeed a supergiant companion. A point-like NIR source is located at the position of the PWN \dieci, although its morphology suggests a foreground star despite its positional coincidence with the nebula. The photometry of the counterpart to the unclassified source \due indicates it is an LMXB with a giant companion star, possibly a SyXB.

We also presented optical/NIR observations of the two LMXBs \cinque and \sette, searching for a counterpart based on their \xmm position. We detected only one source compatible with the position of \cinque, whose magnitude is consistent with what is expected for an LMXB. This  supports its association with the X-ray source. Twelve NIR candidates are consistent with \sette: a \chan position of the source is necessary to select the counterpart. 
Table~\ref{results} summarizes the results of our counterpart search in comparison with previous results. 

\begin{table*}
\caption{~Results of our optical/NIR counterpart search. The upper part of the table lists sources that we observed with \chan.  For the last two sources we obtained \xmm positions from the literature. Candidate counterparts previously proposed are discarded or confirmed based on the \chan position (see text). The coordinates of the counterparts in the table are from 2MASS when available, from our astrometry elsewhere.}
\label{results}
\begin{tabular}{l@{\,}l@{\,}l@{\,}l@{\,}l@{\,}l}
\hline
   Source & Classification & Counterparts & New        & \ra(J2000) & \dec(J2000)  \\
          &                & in literature & counterpart &  & \\
\hline
& & & & & \\
\chan & & & & & \\
\hline
  \uno & SyXB & M2 III$^{\dag(a)}$ & {\it confirmed} & $16^{\rmn{h}}$~$19^{\rmn{m}}$~$33\fs348^{2MASS}$ & $-28\degr$~$07\arcmin$~$39\farcs 74^{2MASS}$ \\
  \due & LMXB$^{1}$ & none & yes: K, M or G (unlikely) III & $16^{\rmn{h}}$~$29^{\rmn{m}}$~$12\fs9$~$\pm~0\farcs 1$  & $-46\degr$~$02\arcmin$~$50\farcs 58$ $\pm~0\farcs 1$  \\
  \tre & SFXT~(eclipsing) & O/B I star$^{\dag(b)}$ & {\it confirmed} & $16^{\rmn{h}}$~$48^{\rmn{m}}$~$06\fs56^{2MASS}$ & ~$-45\degr$~$12\arcmin$~$06\farcs 8^{2MASS}$ \\
  \quattro & CV & dwarf nova$^{\dag(c)}$ & {\it confirmed} & $16^{\rmn{h}}$~$49^{\rmn{m}}$~$55\fs64^{2MASS}$ & $-33\degr$~$07\arcmin$~$02\farcs 1^{2MASS}$ \\
  \sei & obscured HMXB & in $^{(d)}$:~{\it excluded} & 2MASS J17155645-3851537 & $17^{\rmn{h}}$~$15^{\rmn{m}}$~$56\fs46^{2MASS}$ & ~$-38\degr$~$51\arcmin$~$53\farcs 7^{2MASS}$ \\
  \otto & SFXT & none & 2MASS~J17430133-3622221 &  $17^{\rmn{h}}$~$43^{\rmn{m}}$~$01\fs34^{2MASS}$ & $-36\degr$~$22\arcmin$~$22\farcs 2^{2MASS}$\\
  \nove & NS LMXB (dipper) & 6 NIR candidates$^{(e)}$ & Select candidate 1 (see text) & $17^{\rmn{h}}$~$59^{\rmn{m}}$~$45\fs5$~$\pm 0.1\farcs$ & $22\degr$~$01\arcmin$~$39\farcs 6$~$\pm 0.1\farcs$  \\ 
  \dieci & PWN & in$^{(f)}$:~{\it excluded} & yes/tentative & $18^{\rmn{h}}$~$49^{\rmn{m}}$~$01\fs553$~$\pm 0.1\farcs$ & $-00\degr$~$01\arcmin$~$17\farcs 20$~$\pm 0.1\farcs$\\
  \undici & L/IMXB$^{1}$ & F8 star$^{\dag(g)}$ & {\it confirmed} &  $19^{\rmn{h}}$~$30^{\rmn{m}}$~$50\fs76^{2MASS}$ & $+05\degr$~$30\arcmin$~$58\farcs 3^{2MASS}$   \\
\hline
& & & & & \\
\xmm & & & &  \\
\hline
\cinque & NS LMXB (eclipsing) & none & yes & $17^{\rmn{h}}$~$10^{\rmn{m}}$~$12\fs6$~$\pm 0.1\farcs$ & $-28\degr$~$07\arcmin$~$51\farcs 0$~$\pm 0.1\farcs$ \\
\sette & UCXB & none & 12 candidates (see text) & - &  - \\
\hline
\end{tabular}
\begin{flushleft}
{\footnotesize $^{1}$: New classification}

{\footnotesize $^{\dag}$: Optical/NIR spectrum reported in the literature.}

{\footnotesize $^{(a)}$~USNO-A2.0~U0600\_20227091: \citet{2007A&A...470..331M}. $^{(b)}$~2MASS~J16480656-4512068;~\citet{2005ATel..599....1K};~\citealt{2006A&A...453..133W};~\citet{2008A&A...484..783C}. $^{(c)}$~USNO~A2-0~U0525-24170526:\citet{2008A&A...482..113M}.  $^{(d)}$~\citet{2005A&A...432L..49S}, also indicates the presence of several further NIR objects. $^{(e)}~$~\citet{2008A&A...484..783C}. $^{(f)}$~2MASS~J18490182-0001190: \citet{2008A&A...482..731R}\\}
\end{flushleft}
\end{table*}
\newpage

\newpage

\bibliographystyle{mn} \bibliography{h1743-bib.bib}

\end{document}